\documentclass[a4paper,3p,number,sort&compress]{elsarticle}

\usepackage[utf8]{inputenc}
\usepackage{graphics,epsfig,graphicx,float,caption,subcaption,color}
\usepackage{amsmath,amssymb,amsbsy,amsfonts,amsthm}
\usepackage{url}
\usepackage{boxedminipage}
\usepackage{algorithm}
\usepackage{algpseudocode}
\usepackage{bm}
\usepackage{mathabx}

\usepackage[sf,bf,tiny]{titlesec}
\usepackage[plainpages=false, colorlinks=true,
   citecolor=blue, filecolor=blue, linkcolor=blue,
   urlcolor=blue,backref=true]{hyperref}
\usepackage{enumitem}

\newcommand{\bs}{\boldsymbol}

\usepackage{cleveref,svg} 
\usepackage[toc,page]{appendix}

\begin{document}

\begin{frontmatter}

\title{A Fourier spectral immersed boundary method with exact translation invariance, improved boundary resolution, and a divergence-free velocity field}

\author[1]{Zhe Chen\corref{cor1}}
\cortext[cor1]{Corresponding author.}
\ead{zc1291@cims.nyu.edu}
\author[1]{Charles S. Peskin}

\address[1]{Courant Institute of Mathematical Sciences, New York University, 251 Mercer Street, New York, NY, 10012, USA}

\begin{abstract}
This paper introduces a new immersed boundary (IB) method for viscous incompressible flow, based on a Fourier spectral method for the fluid solver and on the nonuniform fast Fourier transform (NUFFT) algorithm for coupling the fluid with the immersed boundary. The new Fourier spectral immersed boundary (FSIB) method gives improved boundary resolution in comparison to the standard IB method. The interpolated velocity field, in which the boundary moves, is analytically divergence-free. 
The FSIB method is gridless and has the meritorious properties of volume conservation, exact translation invariance, conservation of momentum, and conservation of energy. We verify these advantages of the FSIB method numerically both for the Stokes equations and for the Navier-Stokes equations in both two and three space dimensions. The FSIB method converges faster than the IB method. In particular, we observe second-order convergence in various problems for the Navier-Stokes equations in three dimensions. The FSIB method is also computationally efficient with complexity of $O(N^3\log(N))$ per time step for $N^3$ Fourier modes in three dimensions.

\end{abstract}


\end{frontmatter}

\section{Introduction}

The immersed boundary (IB) method is generally applicable to problems
 of fluid-structure interaction \citep{peskinImmersedBoundaryMethod2002}.  The immersed boundary can be thin, and
 when idealized as infinitely thin it applies a singular force density
 (i.e., a delta-function layer of force per unit volume) to the
 surrounding fluid.  An example of such a thin elastic boundary immersed in a viscous incompressible fluid is a
 heart valve leaflet, and this was the motivating example for introduction of the IB method \citep{peskinNumericalAnalysisBlood1977}.  The IB method smooths the
 singularity by introducing a kernel $\delta_h^{(n)}$ which is a smoothed
 approximation to the Dirac delta function and is finitely supported on a box, the size of which in each space direction is some small integer $n$ times the meshwidth $h$. This smoothing, which
 affects not only the application of force to the fluid but also the
 evaluation of the fluid velocity field at the location of the
 immersed boundary, is a source of inaccuracy since it gives the
 computational immersed boundary an effective thickness, on the order
 of a meshwidth, that is not part of the mathematical formulation
 of the problem but is needed for computational reasons.

Moreover, the interpolated velocity field of the standard IB
 method does not have zero divergence, even when the grid velocity
 field that is being interpolated is discretely divergence-free.  A
 consequence of this is a systematic volume leak that is an especially
 disturbing kind of numerical error in some applications, even though
 it can be made arbitrarily small by refinement of numerical
 parameters. There are several ways to reduce the volume leak by a large constant factor with extra effort, e.g. a modified finite difference operator \citep{peskinImprovedVolumeConservation1993} or the MAC discretization \citep{griffithVolumeConservationImmersed2012}. A more recent development is a divergence-free IB method that involves a vector potential computed on a staggered grid, interpolated as in the standard IB method (but with a delta-function kernel that
 has three continuous derivatives), and then differentiated to obtain a divergence-free velocity field in which the immersed boundary moves \citep{baoImmersedBoundaryMethod2017}. The present paper has in common with \citep{baoImmersedBoundaryMethod2017} that the volume leak is eliminated by making the interpolated velocity field continuously divergence-free. This happens here in a much more natural way, however, since our pseudo-spectral fluid solver works directly with a
 continuously divergence-free representation of the fluid velocity
 field, and since our interpolation method is equivalent to the
 direct evaluation of that velocity field at the immersed boundary. Neither the vector potential nor the staggered grid is needed in our new method.

Another issue with the immersed boundary method is translation
 invariance. On a periodic (or infinite) domain, a computational mesh of a given
 meshwidth can be shifted by an arbitrary amount (not necessarily an
 integer multiple of a meshwidth) in each coordinate direction.  It
 would be ideal for the computed solution to be independent of any
 such shift, except for being sampled on the shifted grid.  For this
 to happen, it must be the case that the influence of one immersed
 boundary point on another, even though that influence occurs through
 the fluid grid, should depend only on the vector that connects the
 two points, and not on how those two points are situated in relation
 to the fluid grid. In the standard IB method \citep{peskinImmersedBoundaryMethod2002}, the issue of translation invariance is addressed by introducing a sum-of-squares condition for the delta-function kernel that is used in the velocity interpolation
 and the force spreading.  The sum-of-squares condition ensures
 translation invariance for the self-interaction (mediated by the
 fluid grid) of any immersed boundary point, and it also ensures that
 pairwise interactions are bounded by the translation-invariant 
 self-interaction. Recently, the translation invariance for pair interaction has been improved through the introduction of
 Gaussian-like delta-function kernels with bounded support \citep{baoGaussianLikeImmersedBoundary2016}. Exact translation invariance has not been achieved in the context of
 immersed boundary methods, and indeed it can be shown that exact
 translation invariance is incompatible with bounded support of the
 regularized delta functions.  Although regularized delta functions
 with unbounded support exist, their use would seem to be impractical,
 since the cost of interpolation and force spreading would then be
 proportional to the product of the number of immersed boundary points
 and the number of fluid grid points.  Thus, it might well be believed
 that exact translation invariance is unachievable within the
 framework of the IB method.

The purpose of this article is to overcome these limitations of the IB method. We introduce a new Fourier spectral immersed boundary (FSIB) method for a thin elastic boundary immersed in a viscous incompressible fluid. The fluid equations, including both the Stokes equations and the Navier-Stokes equations, are solved by a Fourier spectral method,
 also known as a pseudospectral method \citep{gottliebNumericalAnalysisSpectral1977a}. The force spreading and the velocity interpolation steps of the IB method
 are here expressed in terms of finite Fourier series. The coefficients of the Fourier series can be calculated efficiently by a fast algorithm, the Non-uniform Fast Fourier Transform (NUFFT) \citep{duttFastFourierTransforms1993}. This plays the role of the force spreading step of the IB method. The Fourier series can be evaluated at any location in physical space by the NUFFT as well, and this plays the role of the velocity interpolation step of the IB method. We use a fast and parallel implementation of the NUFFT called the finufft \citep{barnettParallelNonuniformFast2019,barnettAliasingErrorExp2020}. From an algorithmic point of view, there is no need for
 any regularized delta function and the FSIB method does not employ an Eulerian grid in the physical space. This is the key to translation invariance. 
 
 Despite the foregoing, we prove herein that
 our new method is equivalent to an IB method with a 
 `$\mathtt{sinc}$' function kernel as the regularized delta function. Moreover, the FSIB method has the duality of the force spreading and the velocity interpolation that is similar to the IB method, and the `$\mathtt{sinc}$' kernel even satisfies the conditions of the standard IB kernels continuously and discretely. Thus, the FSIB method preserves the merits of the IB method including the conservation of momentum and the conservation of energy. 
 
 Related to the present paper but in a different field is \citep{mitchellEfficientFourierBasis2019}, in which the Vlasov-Poisson equations are solved by a Fourier spectral method. In this reference, however, the counterpart of the immersed boundary is a collection of point chargers that are smoothed by a Gaussian shape function. In contrast to this, our motivation in the present paper is to avoid any such smoothing, and of course, the Vlasov-Poisson equations are different from the Navier-Stokes equations.

The structure of the present paper is as follows. In \cref{sec:equations}, we describe the continuum formulation that is the foundation of the FSIB method, and in \cref{sec:discretization}, we discuss the
 details of discretization. In \cref{sec:relation}, we show that the FSIB method can be viewed as an
 immersed boundary method with a `$\mathtt{sinc}$' function kernel, which is globally supported, as the
 regularized delta function. We will also show that the FSIB method has exact translation invariance,
 and its time-continuous but spatially discretized version
 conserves both energy and momentum. In \cref{sec:numerical}, a series of numerical experiments will be provided to verify these properties of FSIB in Stokes flow and in Navier-Stokes flow in two space dimensions and also in three space dimensions. In this section, we also study the convergence rate and boundary resolution of the FSIB method.

\section{Mathematical Formulation of the FSIB method}\label{sec:equations}

In this section, we provide a general description of the equations used in the FSIB method. We consider a thin, massless, elastic boundary $\Gamma$, immersed in a viscous incompressible fluid. The fluid is described in Eulerian form by the Navier-Stokes equations.

\begin{equation}\label{eq:navierstokes}
\left\{\begin{array}{l}
\rho\left(\frac{\partial \boldsymbol{u}}{\partial t}+\boldsymbol{u} \cdot \nabla \boldsymbol{u}\right)+\nabla p=\mu \Delta \boldsymbol{u}+\boldsymbol{f}, \\
\nabla \cdot \boldsymbol{u}=0,
\end{array}\right.
\end{equation}
where $\bs u(\bs x,t)$ is the fluid velocity, $p(\bs x,t)$ is the fluid pressure, and
 where $\bs f(\bs x,t)$ is the force per unit volume applied to the fluid.  Our principal use of $\bs f(\bs x,t)$ will be to represent the force per unit
 volume applied to the fluid by the immersed elastic boundary, in
 which case $\bs f(\bs x,t)$ will be a delta-function layer with support on the
 immersed boundary, but sometimes we will also consider given external
 forces that drive a flow. The constant parameters $\rho$ and $\mu$ in
 equation \cref{eq:navierstokes} are the mass density and the dynamic viscosity of the
 fluid, respectively.

The spatial domain occupied by the fluid will be 3-torus, that is, a
 cube with periodic boundary conditions.  It is equivalent, however,
 to say that the fluid occupies all of $\mathcal{R}^3$ and that the functions $\bs u$,
 $p$, and $\bs f$ are periodic with some specified period $L$ in all three
 spatial dimensions. Both points of view will be useful herein.
 We will sometimes consider the two-dimensional case in numerical experiments, 
 but our description of the method will be three-dimensional.

As in the IB method, the immersed boundary and the fluid are coupled as follows

\begin{equation}\label{eq:force_spreading}
\left\{\begin{array}{l}
\boldsymbol{f}(\boldsymbol{x}, t)=\int \boldsymbol{F}(\boldsymbol{\theta}, t) \delta(\boldsymbol{x}-\boldsymbol{X}(\boldsymbol{\theta}, t)d \boldsymbol{\theta},\\
\bs U(\bs \theta , t) = \frac{\partial \boldsymbol{X}}{\partial t}(\boldsymbol{\theta}, t)=\boldsymbol{u}(\boldsymbol{X}(\boldsymbol{\theta}, t), t)=\int \boldsymbol{u}(\boldsymbol{x}, t) \delta(\boldsymbol{x}-\boldsymbol{X}(\boldsymbol{\theta}, t)) d \boldsymbol{x},
\end{array}\right.
\end{equation}
where $\boldsymbol{F}(\boldsymbol \theta,t )$ denotes the force density exerted on the fluid by the immersed boundary, $\boldsymbol{X}(\boldsymbol \theta,t)$ denotes the location of the immersed boundary and $\bs U (\bs \theta,t)$ denotes the velocity of the immersed boundary, all parameterized by the Lagrangian variable $\boldsymbol{\theta}$. The force density described on the immersed boundary is coupled to the fluid through convolution with the Dirac delta function, and the velocity of the immersed boundary is similarly obtained by convolution of the velocity of the fluid with the Dirac delta function.

The foundation of the FSIB method is a rewrite of the above equations in terms of Fourier series. Any periodic function $g \in \mathcal L^2_{\Omega_L}$ on the periodic box $\Omega_L = [0, L]^3$ can be written as:

\begin{equation}\label{eq:contfourseri}
    g(\boldsymbol x)=\sum_{\boldsymbol k \in \mathcal{K}} \hat{g}(\boldsymbol k) \exp (i \boldsymbol k \cdot \boldsymbol x),
\end{equation}
where the $\hat{g}(\boldsymbol k)$ denotes the Fourier series coefficient of wavenumber $\boldsymbol k \in \mathcal{K}$ and $\mathcal{K}=\frac{2\pi}{L}\mathbb{Z}^3$. We denote 
\begin{equation}\label{eq:hat_operator}
\hat{g} = \mathcal{H}(g),
\end{equation} 
where the operator $\mathcal{H}$ is the map from a periodic function to its Fourier coefficients and it gets the name `H' because it puts a hat on the function. In this way, all of the periodic functions $\bs f, p$ and $\bs u$ can be written as Fourier series with coefficients $\hat{\bs f}, \hat{p}$ and $\hat{\bs u}$. In the following, we omit arguments such as $t$, $\bs x$, $\bs k$, and $\bs \theta$ in the notations for convenience as long as the meaning is clear. To express the Navier-Stokes \cref{eq:navierstokes} in the Fourier domain, we first note that pressure $p$ satisfies the pressure Poisson equation:

\begin{equation}\label{eq:pressure_poisson}
\Delta p = \nabla \cdot (\bs f - \rho \bs u\cdot \nabla \bs u),
\end{equation}
which is obtained from taking the divergence of momentum equation of Navier-Stokes equations (\ref{eq:navierstokes}) and eliminating the divergence-free terms. 

We denote the nonlinear term $\bs s(\bs u):=\bs u \cdot \nabla \bs u$ and its Fourier coefficients as $\hat{\bs s}(\hat{\bs u})$. The calculation of $\bs s$ usually takes place in the physical space as

\begin{equation}\label{eq:nonlinear_physical}
\hat{\bs s}(\hat{\bs u}) = \mathcal{H}\left( (\mathcal{H}^{-1}(\hat{\bs u}))\cdot (\mathcal{H}^{-1}(i\bs k \hat{\bs u}))\right).
\end{equation}
In Fourier space, the pressure Poisson equation becomes

\begin{equation}\label{eq:pressure_poisson_k}
\hat{p}=-\frac{i\bs k \cdot (\hat{\bs f}-\rho \hat{\bs s}(\hat{\bs u}))}{|\bs k|^2}.
\end{equation}
The case $\bs k=0$ for the denominator makes no difficulty because the addition of a constant to the pressure makes no difference, so we can set $\hat{p}(\bs k=0)$ as any constant.

With the use of \cref{eq:pressure_poisson_k}, the Fourier series version of \cref{eq:navierstokes} becomes 

\begin{equation}\label{eq:contfoursoln}
\frac{\partial \hat{\bm{u}}(\boldsymbol k)}{\partial t}=(\bm{I}-\frac{\bm{k} \bm{k}^T}{|\bm{k}|^2})(\frac{\hat{\bm{f}}(\bm k)}{\rho}-\hat{\bs s}(\hat{\bs u}, \bm k))-\frac{\mu}{\rho}|\bm{k}|^2\hat{\bm{u}}(\bm k),
\end{equation}
where $\bs k \in \mathcal{K}$. Note that $\bm{I}-\frac{\bm{k} \bm{k}^T}{|\bm{k}|^2}$ is the projection operator onto the divergence-free space in the Fourier domain. Inherently, this enforces the divergence-free condition of velocity, i.e. $\bs k \cdot \hat{\bs u}=0$ in Fourier space or $\nabla \cdot \bs u=0$ in physical space, analytically.

By Fourier series, the coupling between the immersed boundary and the fluid in \cref{eq:force_spreading} can be evaluated as

\begin{equation}\label{eq:force_spreading_1}
\hat{\bs f}(\bs k)=\frac{1}{L^3} \int_{\bs x \in \Omega_L} \bs f(\bs x) \exp(-i\bs k \cdot \bs x) d\bs x =  \frac{1}{L^3} \int_{\bs \theta \in \Gamma} \bs F(\bs \theta) \exp(-i \bs k \cdot \bs X(\bs \theta)) d \bs \theta,\quad \bs k \in \mathcal{K},
\end{equation}

\begin{equation}\label{eq:velocity_interpolation_1}
     \bm U(\bm \theta)= \bm u(\bm X(\bm \theta))=\sum_{\bm{k} \in \mathcal{K}}\hat{u}(\bm{k})\exp(i\bm{k}\cdot \bm{X}(\bm \theta)).
\end{equation}
 It is encouraging that the singularity of the Dirac delta function no longer appears in these equations.
 
In summary, for a given system state with Lagrangian coordinates of the immersed boundary in physical space $\bs X(\bs \theta,t)$ and fluid velocity in Fourier space $\hat{\bs u}(\bs k,t)$, we first compute the force in Fourier space $\hat{f}(\bs k,t)$ from \cref{eq:force_spreading_1}. Then by evaluating the right-hand side of \cref{eq:contfoursoln}, we are able to obtain the rate of change of the velocity $\hat{\bs u}(\bs k,t)$. By evaluating the Fourier series \cref{eq:velocity_interpolation_1}, we are able to obtain the velocity of the immersed boundary $\bs U(\bs \theta,t)$. Last, we are able to compute the rate of change of the location of the immersed boundary $\bs X(\bs \theta,t)$ as $\partial \bs X(\bs \theta,t) /\partial t=\bs U(\bs \theta,t)$. 

\section{Numerical implementation of the FSIB method}\label{sec:discretization}

First, we provide some foresight into the framework of this section. According to the Fourier spectral solver in \cref{eq:contfoursoln}, we need to compute $\hat{\bs f}$ and $\hat{\bs s}(\hat{\bs u})$ for given $\hat{\bs u}$ and $\bs X$, all of which are discussed in \cref{sec:spatial_discretization}. Note that $\hat{\bs f}$ is acquired by \cref{eq:force_spreading_1} and \cref{eq:velocity_interpolation_1} is used to update the velocity of the immersed boundary. Both are types of NUFFT and are shown in detail in appendix \ref{appendix:nufft}. Furthermore, a temporal integrator is needed to evolve the \cref{eq:contfoursoln} and we show this process in \cref{sec:temporal_discretization}.  We will summarize the FSIB algorithm in \cref{sec:algo_summary}. 

It should be noticed that the Lagrangian variables including $\bs X$, $\bs F$ and $\bs U$ are defined in physical space while the Eulerian variables like $\hat{\bs u}$ are defined in Fourier space. These two types of variables are connected by force spreading and velocity interpolation. Note that the FSIB method has no Eulerian grid in the physical space. This is unlike the standard IB method, and is one of the best features of the FSIB method.

\subsection{Spatial approximation}\label{sec:spatial_discretization}

In the standard IB method, force spreading and velocity interpolation are the most important steps, as they connect Eulerian variables and Lagrangian variables. The singular Dirac delta function in  \cref{eq:force_spreading} is approximated by a regularized delta function that has a finitely large height and finitely small width \citep{peskinImmersedBoundaryMethod2002}. It thus computes the force spreading and the velocity interpolation by convolutions with the regularized kernel in physical space. In the new FSIB method, however, these two processes are easily done in Fourier space and do not require any regularization of the Dirac delta function in the physical space.
 
In practice, the computation of the coupling between the immersed boundary and the fluid in eq. (\ref{eq:force_spreading_1}) and (\ref{eq:velocity_interpolation_1}) is limited by the number of Lagrangian grid points or wavenumbers of Fourier series. First, it is natural to use a finite number of Fourier modes $\mathcal{K}_{N}$ that is the truncation of $\mathcal{K}$
\begin{equation}\label{eq:kNh}
\mathcal{K}_{N}=\left\{
\begin{array}{ll}
\left\{\frac{2\pi}{L} [-N/2,...,0,...,N/2-1]\right\}^3, & N\text{ is even,} \\
\left\{\frac{2\pi}{L} [-(N-1)/2,...,0,...,(N-1)/2]\right\}^3, & N\text{ is odd,}
\end{array}\right.
\end{equation}
where $N$ is the number of Fourier modes in each dimension so there are $N^3$ Fourier modes in total and the power of $3$ means it is a three-dimensional vector as a Cartesian product. On the truncated range of frequencies $\mathcal{K}_{N}$, we evaluate the integral of \cref{eq:force_spreading_1} with some quadrature $\bs \theta_i$ where $i\in {1,...,N_b}$ and $N_b$ denotes the number of Lagrangian grid points. Note that the Lagrangian grids $\bs \theta_i$ are not necessarily equispaced. This gives the force spreading in the FSIB method as follows

\begin{equation}\label{eq:force_spreading_quad}
\hat{\bs f}(\bs k)=  \frac{1}{L^3} \int_{\bs \theta \in \Gamma} \bs F(\bs \theta) \exp(-i \bs k \cdot \bs X(\bs \theta)) d \bs \theta \approx \frac{1}{L^3} \sum_{j=1}^{N_b}    \omega_j \bs F(\bs \theta_j) \exp(-i \bs k \cdot \bs X(\bs \theta_j)),\quad \bs k \in \mathcal{K}_{N},
\end{equation}
where the $\omega_j$ are quadrature weights. Note that the Lagrangian variable $\theta$ is usually periodic, e.g. a closed curve in two dimensions, in which case a simple equispaced quadrature will give exponential convergence and we also show this numerically in \cref{sec:stokes}.

We compute the velocity interpolation on the Lagrangian grids:

\begin{equation}\label{eq:velo_intp_fini}
\bm U(\bm \theta_j)\approx \sum_{\bm{k} \in \mathcal{K}_{N}}\hat{u}(\bm{k})\exp(i\bm{k}\cdot \bm{X}(\bm \theta_j)),\quad j=1,...,N_b.
\end{equation}

Direct calculations of the new force spreading and the velocity interpolation above (eq. (\ref{eq:force_spreading_quad}) and (\ref{eq:velo_intp_fini})) are quite expensive. Both of them require $O(N_b\cdot N^3)$ for naive for-loops. Moreover, we will show that, in practice, the number of Fourier modes and the Lagrangian grid points are refined in a relationship as $N_b \sim N^2$ in three dimensions, which makes the computational cost $O(N^5)$. This is unaffordable and also why we need the NUFFT algorithm to reduce the computational cost to $O(N^3\log(N))$. The force spreading in \cref{eq:force_spreading_quad} is from nonuniform grids to uniform grids and is a type-1 NUFFT. The velocity interpolation in \cref{eq:velo_intp_fini} is from uniform grids to nonuniform grids and is a type-2 NUFFT. In order not to break the smooth flow of the deduction process, we leave the details of using the type-1 and the type-2 NUFFT to compute the force spreading and the velocity interpolation to appendix \ref{appendix:nufft}. For now, we assume $\hat{\bs f}(\bs k)$ in \cref{eq:force_spreading_quad} and $\bs U(\bs \theta)$ in \cref{eq:velo_intp_fini} are computed by an efficient algorithm, i.e. NUFFT, with the complexity of $O(N^3\log(N))$.

Another term that is tricky to compute is the nonlinear term $\bs s(\bs u) = \bs u \cdot \nabla \bs u$ in Fourier space. By the convolution theorem, the dot product in physical space becomes convolution in Fourier space. The multiplication in physical space $\bs u \cdot \nabla \bs u$ has only $O(N^3)$ complexity, but the convolution in Fourier space has $O(N^6)$ complexity, which is again unaffordable. Thus, it is better to compute the nonlinear term $\bs s(\bs u)$ by multiplication in physical space. An exact computation of \cref{eq:nonlinear_physical} in $\mathcal{K}_N$ is 

\begin{equation}\label{eq:nonlinear_kspace}
\hat{\bs s}(\hat{\bs u}) = \mathcal{T}_{N} \circ \mathcal{H}\left( (\mathcal{H}^{-1}(\hat{\bs u}))\cdot (\mathcal{H}^{-1}(i\bs k \hat{\bs u}))\right),
\end{equation}
where $\mathcal{T}_{N}$ denotes the truncation operator on $\mathcal{K}_N$. 

A naive way to compute $\hat{\bs s}(\hat{\bs u})$ is the following. For input $\hat{\bs u}$, we first compute $\widehat{\nabla \bs u}=i\bs k\cdot \hat{\bm{u}}$ in Fourier space. Then convert back to physical space to get $\bs u$ and $\nabla \bs u$ by inverse FFT and do the multiplication to get $\bs s(\bs u)= \bs u \cdot \nabla \bs u$. Then convert $\bs s(\bs u)$ back into Fourier space by FFT to get $\hat{\bs s}(\hat{\bs u})$. A complication here is that the multiplication in physical space causes aliasing error because it is a multiplication of two bandlimited functions $\bs u$ and $\nabla \bs u$, but in Fourier space, the convolution of two functions that are bandlimited in $\mathcal{K}_{N}$ will contribute to nonzero Fourier coefficients that are outside of $\mathcal{K}_{N}$. Since we only resolve $\bs s(\bs u)$ on Fourier modes $\mathcal{K}_{N}$ by FFT, those frequencies outside of $\mathcal{K}_{N}$ will pollute low frequencies inside of $\mathcal{K}_{N}$, which is called aliasing. Moreover, such aliasing error will make the Fourier spectral method suffer from numerical instability \citep{goodmanStabilityUnsmoothedFourierMethod1994, phillips1959example}. 

The famous ``three-halves" rule, also known as Orszag's ``two-thirds" rule \citep[Chapter~11]{boydChebyshevFourierSpectral}, is applied to prevent such aliasing. First, we know the two Fourier coefficients $\hat{\bs u}$ and $\widehat{\nabla \bs u}=i \bs k \hat{\bs u}$ that have a length of $N$ on each dimension. A centered zero padding is implemented on each space dimension to length $\Tilde{N}>\frac{3}{2}N$ such that the range of frequencies is now $\mathcal{K}_{\tilde{N}}$. Then, an inverse FFT will give $\bs u$ and $\nabla \bs u$ in physical space of length $\Tilde{N}$ on each dimension. We then compute the multiplication to get $\bs s(\bs u) = \bs u \cdot \nabla \bs u$. Finally, one FFT gives $\widehat{\bs u\cdot \nabla \bs u}$ and truncate the frequency range from $\mathcal{K}_{\tilde{N}}$ to $\mathcal{K}_{N}$, i.e. $\hat{\bs s}(\hat{\bs u})$ for $\bs k \in \mathcal{K}_{N}$. As long as $\Tilde{N}>\frac{3}{2}N$, it can be proved that the frequencies in $\mathcal{K}_{N}$ are not polluted by the high-frequency modes outside of $\mathcal{K}_{N}$ and there is no aliasing error for the Fourier modes in $\mathcal{K_N}$. This is why it is called the ``three-halves" rule. 

In general, the ``three-halves'' rule provides us a way to compute the multiplication of two bandlimited functions efficiently in Fourier space with complexity $O(N^3\log(N))$ without aliasing error. One may disregard the details here and just consider the ``three-halves'' rule as a black-box algorithm to compute \cref{eq:nonlinear_kspace} exactly.

Until now, we have obtained all the terms that are needed in the Fourier spectral method, so we can compute the rate of change of the velocity numerically below, which is a spacial discretization and Fourier truncation from \cref{eq:contfoursoln}.

\begin{equation}\label{eq:fourier_spectral_sol}
\frac{\partial \hat{\bm{u}} (\bm k)}{\partial t}=(\bm{I}-\frac{\bm{k} \bm{k}^T}{|\bm{k}|^2})(\frac{\hat{\bm{f}}(\bm k)}{\rho}-\hat{\bs s}(\bm k))-\frac{\mu}{\rho}|\bm{k}|^2\hat{\bm{u}}(\bm k):=\hat{V}(\hat{\bm{f}},\hat{\bm{u}})-\frac{\mu}{\rho}|\bm{k}|^2\hat{\bm{u}}(\bm k),\quad \bm k \in \mathcal{K}_{N},
\end{equation}
where we denote the first term as $\hat{V}(\hat{\bs f},\hat{\bs u})$ for the benefit of describing the temporal integrator in \cref{sec:temporal_discretization}. 

Note that $|\bm{k}|^2$ is in the denominator in the projection operator $\bm{I}-\frac{\bm{k} \bm{k}^T}{|\bm{k}|^2}$ and we need to avoid the divide-by-zero issue. In the code, the $|\bs k|^2$ is a three-dimensional matrix and we just need to artificially set $|\bs k|^2=1$ when $|\bs k|=0$, which should be the first entry of the matrix, and everything follows naturally as matrix operations. This is because this term comes from the pressure Poisson equation and the pressure $p$ makes no difference by adding any constant which has been discussed in \cref{eq:pressure_poisson_k}.

\subsection{Temporal approximation}\label{sec:temporal_discretization}

We denote the time step as $\Delta t$. Thus, for a given time period $[0,T]$, we have the temporal grids $t_m=m \Delta t$, where $m=0,...,N_t-1,N_t$ and $\Delta t=T/N_t$. 

Note that the FSIB method is gridless and the system should be initialized by given $\bs \hat{u}^0$ and $\bs X^0$. The velocity of the immersed boundary $\bs U^0$ is easily obtained by a Fourier series evaluation which is the velocity interpolation with NUFFT. Now we can assume, at any time point $t_m$, $m=0,...,N_t-1$, the variables $\hat{\bs u}^m$, $\bs U^m$ and $\bs X^m$ are known inputs to the temporal integrator. Similar to the standard IB method, a Runge-Kutta multistep temporal integral is used here to solve \cref{eq:fourier_spectral_sol}. First, a half-time step is carried out explicitly to update the location of the immersed boundary $\bm{X}^{m+1/2}$ in \cref{eq:spatial_1}. Using this geometry, the Force density $\bm{F}^{m+1/2}$ is obtained, usually by elasticity. Thus, $\hat{\bm{f}}^{m+1/2}$ is easily available through the force spreading in \cref{eq:force_spreading_quad} with the NUFFT. 

For improved numerical stability, the diffusion term will be computed implicitly. As in the standard immersed boundary method, we first use the backward Euler for a half time step with inputs $\hat{\bm{f}}^{m+1/2}$ and $\hat{\bm{u}}^m$ to get $\hat{\bm{u}}^{m+1/2}$ in \cref{eq:spatial_2}. Then, for inputs $\hat{\bm{u}}^{m+1/2}$ and $\hat{\bs f}^{m+1/2}$, we use the Crank-Nicolson method for a whole time step to solve for $\hat{\bm{u}}^{m+1}$ in \cref{eq:spatial_3}. Using velocity interpolation in \cref{eq:velo_intp_fini}, the velocity on the Lagrangian grids $\bs U^{m+1/2}$ and $\bs U^{m+1}$ are easily computed. The location of the immersed boundary at the next time step $X^{m+1}$ is then updated in \cref{eq:spatial_4}.

\begin{flalign}
&\bs X^{m+1/2}_\theta=\bs X_{\bs \theta}^m+\bs U^m\frac{\Delta t}{2} \label{eq:spatial_1}\\
&\frac{\hat{\bs u}^{m+1/2}-\hat{\bs u}^m}{\Delta t/2}=\hat{V} (\hat{\bs f}^{m+1/2},\hat{\bs u}^m)-\frac{\mu|\bs k|^2}{\rho} \hat{\bs u}^{m+1/2} \label{eq:spatial_2} \\
&\frac{\hat{\bs u}^{m+1}-\hat{\bs u}^m}{\Delta t}= \hat{V} (\hat{\bs f}^{m+1/2},\hat{\bs u}^{m+1/2})-\frac{\mu|\bs k|^2(\hat{\bs u}^m+\hat{\bs u}^{m+1})}{2\rho} \label{eq:spatial_3} \\
&\bs X^{m+1}_{\bs \theta}=\bs X_{\bs \theta}^m+\bs U^{m+1/2} \Delta t \label{eq:spatial_4}
\end{flalign}

\subsection{Algorithm summary}\label{sec:algo_summary}

Up to now, we have introduced all processes of the FSIB algorithm and we summarize the algorithm in this section. The overall framework of the FSIB method is stated in \cref{alg:FSIB}. First, the system is initialized by $\hat{\bs u}^0$, $\bs X^0$, and $\bs U^0$. Note that the velocity of the immersed boundary $\bs U^0$ is then obtained by the velocity interpolation with a type-2 NUFFT from $\hat{\bs u}^0$ as shown in \cref{eq:velo_intp_fini}. In the temporal for-loop, everything is computed as matrix operations. Note that the implicit methods including the backward Euler and the Crank-Nicolson become diagonal in Fourier space. Thus they are implemented by simple divisions in Fourier space.

\begin{algorithm}
\caption{The algorithm framework of FSIB method}\label{alg:FSIB}
\begin{algorithmic}
\Require $\hat{\bs u}^0$, $\bs X^0$ and $\bs U^0$ \Comment{Initialization}
\For {$m=0,...,N_t-1$}
\State $\bs X^{m+1/2}\gets \bs X^{m}+\frac{\bs U^m \Delta t}{2}$
\State Compute $\bs F^{m+1/2}$ from $\bs X^{m+1/2}$ \Comment{e.g. potential force or target points}
\State Compute $\hat{\bs f}^{m+1/2}$ from $\bs F^{m+1/2}$  \Comment{Type-1 NUFFT as \cref{eq:force_spreading_quad}}
\State $\hat{\bs u}^{m+1/2}\gets (\frac{\Delta t}{2}\hat{V} (\hat{\bs f}^{m+1/2},\hat{\bs u}^m)+\hat{\bs u}^{m})/(1+\frac{\mu \Delta t |\bs k|^2}{2\rho})$ \Comment{\cref{eq:spatial_2}}
\State $\hat{\bs u}^{m+1}= (\Delta t \hat{V} (\hat{\bs f}^{m+1/2},\hat{\bs u}^{m+1/2})+(1-\frac{\mu\Delta t |\bs k|^2}{2\rho})\hat{\bs u}^m)/(1+\frac{\mu\Delta t |\bs k|^2}{2\rho})$ \Comment{\cref{eq:spatial_3}}
\State Compute $\bs U^{m+1/2}$ and $\bs U^{m+1}$ from $\bs u^{m+1/2}$ and $\bs u^{m+1}$ \Comment{Type-2 NUFFT as \cref{eq:velo_intp_fini}}
\State $\bs X^{m+1}\gets \bs X^m+\bs U^{m+1/2} \Delta t $
\EndFor
\end{algorithmic}
\end{algorithm}

Moreover, the function $\hat{V}(\hat{\bs f}, \hat{\bs u})$ is called twice at every time step. So, we list the algorithm of $\hat{V}(\hat{\bs f}, \hat{\bs u})$ as \cref{alg:projector}. We first compute the nonlinear term using the ``three-halves" dealiasing rule. Then project $\hat{f}/\rho-\hat{s}$ into divergence-free space by the projection operator. Note that we need to set $|\bs k|^2=1$ when $\bs k=0$. It is worth noticing that all the steps are operated matrix-wised to reach optimal efficiency.

\begin{algorithm}
\caption{Computation of $\hat{V}(\hat{\bs f}, \hat{\bs u})$}\label{alg:projector}
\begin{algorithmic}
\Procedure{$\hat{V}$}{$\hat{\bs f},\hat{\bs u}$}
\State Compute the nonlinear term $\hat{\bs s}(\hat{\bs u})$ as discussed in \cref{sec:spatial_discretization}
\State $P \gets \frac{\hat{\bs f}}{\rho}-\hat{\bs s}(\hat{\bs u})$ \Comment{The vector to be projected}
\State  $Q\gets P-\frac{\bs k \bs k^T}{|\bs k|^2}P$, set $|\bs k|^2=1$ when $\bs k=0$
\State \textbf{return} $Q$
\EndProcedure
\end{algorithmic}
\end{algorithm}

The main calculations include the NUFFT algorithms and matrix operations of the fluid solver, both of which have complexity $O(N^3\log(N))$. Therefore, the overall computational complexity for the FSIB method is $O(N^3\log(N))$ per time step in three dimensions.

\section{Relationship to the standard IB method}\label{sec:relation}

It may appear that the FSIB method is \textit{not} an IB method
 at all, since the hallmark of the IB method is the discretization
 of \cref{eq:force_spreading} through regularization of the Dirac delta function,
 but no such regularization appears explicitly in the FSIB method.
 In this section, however, we will show that the FSIB method
 implicitly does involve such a regularization, which comes from  the
 bandlimited representation of the velocity and force fields that
 appear in \cref{eq:force_spreading_1,eq:velocity_interpolation_1}. This regularization is equivalent
 to the use of a `$\mathtt{sinc}$' function instead of the kernels that are
 normally used in the IB method.  Note that the `$\mathtt{sinc}$' function has
 unbounded support and decays slowly, so a naive implementation based on actual
 use of the `$\mathtt{sinc}$' function would be expensive, but this expense
 is avoided by using the NUFFT algorithm.  Even though the `$\mathtt{sinc}$' function
 is not used explicitly in the FSIB implementation, it is
 instructive to see the close relationship between the standard
 IB method and the FSIB method.  Indeed, as we shall see, the `$\mathtt{sinc}$'
 function satisfies an infinite number of moment conditions of the same
 kind as are satisfied by some standard IB kernels (but the
 standard kernels only satisfy a finite number of these conditions).
 Moreover, the `$\mathtt{sinc}$' kernel satisfies a sum-of-squares condition
 that is also satisfied by standard IB kernels.

 For simplicity, we discuss the relationship between the standard
 IB method and the FSIB method in a one-dimensional context.
 Everything we say extends directly to the two-dimensional and
 to the three-dimensional case, since the kernels used in two
 or in three space dimensions are simply tensor products of
 one-dimensional kernels.

In the standard IB method, the Dirac delta function that appears in \cref{eq:force_spreading} is approximated by regularized delta functions \citep{peskinImmersedBoundaryMethod2002}, which we denote by $\delta^{(n)}_h$. Here $h=L/N$ is the Eulerian mesh width. $N$ is the number of Eulerian grid points in each spatial dimension and we use the same notation as the number of Fourier modes in the FSIB method deliberately because we will demonstrate that these parameters play the same role in the regularization. The integer $n\geq 3$ represents the support of $\delta^{(n)}_h$ in terms of meshwidths. These kernels are constructed for each $n$ from a corresponding function $\phi$ that is independent of $h$ according to the recipe $\delta^{(n)}_h(x)=\phi(x/h)/h$, where $\phi$ is continuous and finitely supported as $\phi(r)=0$ for $|r|\geq n/2$. Note that $\phi$ depends on $n$ for the support and for the conditions (see below) but we leave this dependence understood. Then the functions $\phi(x)$ are uniquely determined by some particular selection of the following conditions depending on whether $n$ is odd or even. 

\begin{equation}\label{eq:tradIBkernels}
\begin{split}
\sum_{j} \phi(r-j)=1,&\quad (\romannumeral 1)\\
\sum_{j \text { even }} \phi(r-j)=\sum_{j \text { odd }} \phi(r-j)=\frac{1}{2},\quad n \text{ even},&\quad (\romannumeral 2) \\
\sum_{j}(r-j)^\alpha \phi(r-j)=0,\quad \alpha = 1,\ldots,n ,&\quad (\romannumeral 3) \\ 
\sum_j \phi^2(r-j)=C,&\quad (\romannumeral 4) \\
\end{split}
\end{equation}
where $C$ is a positive constant to be determined\footnote{The idea of considering this whole family of IB delta functions parameterized by $n$ is due to John Stockie (unpublished communication), who wrote a Maple code to generate any one of them.}.

The purposes of these conditions are discussed in detail in \citep{peskinImmersedBoundaryMethod2002}. In general, the zeroth moment condition (\romannumeral 1) guarantees that the total force spread to the grids remains the same as the total Lagrangian force. The even-odd condition (\romannumeral 2) is stronger than the condition (\romannumeral 1) and avoids the decoupling of two sets of grids consisting of the n-th grid points with n odd or even. The moment condition (\romannumeral 3) enforces the conservation of total torque when $\alpha=1$. Higher order moment conditions are for higher interpolation accuracy order and help determine a kernel uniquely. The sum-of-squares condition (\romannumeral 4) provides an upper estimation for the pair-coupling of the Lagrangian points.  

If $n$ is odd, we use the zeroth moment condition ($\romannumeral 1$), the moment conditions ($\romannumeral 3$) for $\alpha = 1,...,n$ and the sum-of-squares condition ($\romannumeral 4$). If $n$ is even, we use the even-odd condition ($\romannumeral 2$) which implies condition ($\romannumeral 1$), the moment condition ($\romannumeral 3$) for $\alpha=1,...,n$ and the sum-of-squares condition ($\romannumeral 4$). Either case determines a constant $C$ and a unique kernel, which is called the standard n-point kernel $\delta^{(n)}_h$. 

With these kernels, the force spreading and the velocity interpolation approximations are given between the Eulerian grids and the Lagrangian grids as

\begin{equation}\label{eq:force_spreading_IB}
\left\{\begin{array}{l}
f(x_m)=\sum_{j=1}^{N_b} F(\theta_j) \delta_h^{(n)}\left(x_m-X(\theta_j)\right) \Delta \theta,\\
U( \theta_j ) =\sum_{m=1}^{N}u(x_m) \delta_h^{(n)}(x_m-X(\theta_j)) h ,
\end{array}\right.
\end{equation}
where an equispaced Lagrangian grid is used so the quadrature weight is $\Delta \theta$. Unlike the IB method, the FSIB processes these two steps in Fourier space. Nevertheless, it is possible that we convert it into a form like the IB method in physical space. Consider the force spreading \cref{eq:force_spreading_quad} in one dimension and an equispaced Lagrangian grid. So $\omega_j=\Delta \theta$ and, by Fourier series, $f(x_m)=\sum_{k \in \mathcal{K}_N} \hat{f}(k)\exp(i\frac{2\pi}{L}kx)$. Interchange the order of summation and we get

\begin{equation}\label{eq:forc_sprea_eule_FSIB}
f(x_m)=\sum_{j=1}^{N_b} F(\theta_j) \delta_{L,N}\left(x_m-X(\theta_j)\right) \Delta \theta,
\end{equation}
where $\delta_{L,N}(x)=\frac{1}{L}\sum_{k \in \mathcal{K}_N}\exp(i\frac{2\pi}{L}kx)$ is the equivalent kernel in physical space that we are looking for. Similarly, we find the velocity interpolation \cref{eq:velo_intp_fini} is written in physical space as 

\begin{equation}\label{eq:velo_inte_eule_FSIB}
U( \theta_j ) =\sum_{m=1}^{N}u(x_m) \delta_{L,N}(x_m-X(\theta_j)) h.
\end{equation}

Thus, it is concluded the FSIB method has an analog in the IB method using a new kernel $\delta_{L,N}$ because the force spreading \cref{eq:forc_sprea_eule_FSIB} and the velocity interpolation \cref{eq:velo_inte_eule_FSIB} are in the same form with the IB method \cref{eq:force_spreading_IB} but with a different kernel. For simplicity, we now assume that $N$ is odd and thus the FSIB kernel is computed as

\begin{equation}\label{eq:FSIB_kern_odd}
\begin{aligned}
\delta_{L, N}(x) &=\frac{1}{L} \sum_{k=\left(\frac{N-1}{2}\right)}^{\frac{N-1}{2}} e^{i \frac{2 \pi}{L} k x} \\
&=\frac{1}{L} \frac{\sin \left(\frac{N \pi}{L} x\right)}{\sin \left(\frac{\pi}{L} x\right)}.
\end{aligned}
\end{equation}

Now it is easy to see that the FSIB kernel is real, even, and periodic with period $L$. Moreover, we will show that it satisfies the conditions of the standard kernels in \cref{eq:tradIBkernels} except for the odd-even condition (\romannumeral 2) which it does not need to satisfy. First, this continuous kernel has the property that its integral over any one period is equal to $1$. This is obvious from \cref{eq:FSIB_kern_odd} since only the term $k=0$ contributes to the integral. Besides the continuous integral, the discrete zeroth moment also gives 

\begin{equation}
\sum_{j=-\frac{N-1}{2}}^{\frac{N-1}{2}} \delta_{L, N}\left(x-j h\right) \cdot h= \sum_{k=-\frac{N-1}{2}}^{\frac{N-1}{2}} e^{i \frac{2 \pi}{L} k x}\left(\frac{1}{N} \sum_{j=-\frac{N-1}{2}}^{\frac{N-1}{2}} e^{-i \frac{2 \pi}{N} j k}\right) =1,
\end{equation}
since the inner summation is equal to $0$ except for the term $k=0$, and then it is equal to $1$. Note that this identity holds for all real $x$ and thus it is the same as the zeroth condition (\romannumeral 1) of the IB method in \cref{eq:tradIBkernels} after normalization by $h$. The even-odd condition (\romannumeral 2) is not satisfied by the FSIB kernel. This is because the even-odd condition is for the `checkerboard' issue of the Eulerian grid but the FSIB method is gridless, which makes the even-odd condition not necessary.  

An even more remarkable property of $\delta_{L,N}$ related to the translation invariance that is discussed in detail in \cref{sec:trans_invar} is the following:

\begin{equation}\label{eq:tran_inva_kern_FSIB}
\begin{aligned}
&\sum_{j=-\frac{N-1}{2}}^{\frac{N-1}{2}} \delta_{L, N}\left(x_1-j h\right) \delta_{L, N}\left(x_2-j h\right) h \\
&=\frac{1}{L} \sum_{k_1=-\frac{N-1}{2}}^{\frac{N-1}{2}} \sum_{k_2=-\frac{N-1}{2}}^{\frac{N-1}{2}} e^{i \frac{2 \pi}{L}\left(k_1 x_1+k_2 x_2\right)}\left(\frac{1}{N} \sum_{j=-\frac{N-1}{2}}^{\frac{N-1}{2}} e^{-i \frac{2 \pi}{N}\left(k_1+k_2\right) j}\right) \\
&=\sum_{k=-\frac{N-1}{2}}^{\frac{N-1}{2}} e^{i \frac{2 \pi}{L} k\left(x_1-x_2\right)}\\
&=\delta_{L, N}\left(x_1-x_2\right),
\end{aligned}
\end{equation}
where we use the fact that the sum of $j$ on the second line is zero unless $k_1+k_2=0$. This identity essentially means any pair interaction between two points $x_1$ and $x_2$ does not depend on the location of Eulerian grids, which implies translation invariance. The sum-of-squares condition (\romannumeral 4) in \cref{eq:tradIBkernels} of the IB method only provides an upper estimation of the pair-coupling, which is a much weaker condition than the translation invariance. Indeed, we can set $x_1=x_2$ in \cref{eq:tran_inva_kern_FSIB} and get 

\begin{equation}\label{eq:sum_square_FSIB}
\sum_{j=-\frac{N-1}{2}}^{\frac{N-1}{2}} \delta_{L, N}^2\left(x-j h\right) h=\delta_{L, N}(0)=\frac{1}{h},
\end{equation}
which proves the FSIB kernel satisfies the sum-of-squares condition of the IB kernel as well.

It is interesting to see the limit of the period $L$ goes to infinity with a fixed meshwidth $h=L/N$. This is a transition from the periodic boundary condition to the free space and we have

\begin{equation}\label{eq:limit_FSIB_kern}
\lim _{L, N \rightarrow \infty} \delta_{L, N}(x)=\frac{\sin (\pi x / h)}{\pi x}=\frac{1}{h} \frac{\sin (\pi x / h)}{(\pi x / h)}=\frac{1}{h} \mathtt{sinc}\left(\frac{\pi x}{h}\right),
\end{equation}
which we denote as the `$\mathtt{sinc}$' kernel $\delta_h(x)=\frac{1}{h}\mathtt{sinc}(\frac{\pi x}{h})$. This is not a coincidence but points out the essence of the FSIB method. One should notice that the force spreading \cref{eq:forc_sprea_eule_FSIB} and the velocity interpolation \cref{eq:velo_inte_eule_FSIB} are first considered in the form of summation over one period, which means the convolutions with $\delta_{L,N}$ kernel are carried out in one period. One may also consider this in a different way where periodic functions such as the force $f$ or the velocity $u$ are viewed as functions with repeated images in real space $\mathbb{R}$. So the force spreading and the velocity interpolation are convolutions in $\mathbb{R}$ rather than just one period. We know that the truncation of $\mathcal{K}_{N}$ in Fourier space is equivalent to multiplication with a square function as a filter. The square function is centered at $\bs k=0$ and is in the form, no matter whether $N$ is odd or even, as

\begin{equation*}
\tilde{\delta}_h(k)= \left\{
\begin{array}{ll}
1 &\text{, if } -\frac{\pi}{h}\le k< \frac{\pi}{h},  \\
0 &\text{, else,}
\end{array}\right.
\end{equation*}
which we will later show why it is denoted as $\tilde{\delta}_h$. By the convolution theorem, the multiplication in Fourier space is equivalent to a convolution with the inverse Fourier transform in physical space $\mathbb{R}$. We know that the inverse transform of $\tilde{\delta}_h(k)$ is

$$\mathcal{F}^{-1}\circ \tilde{\delta}_h = \frac{1}{2\pi}\int_{-\pi/h}^{\pi/h} \exp(ikx)\,dk=\frac{\sin(x\pi/h)}{x\pi}=\frac{1}{h}\mathtt{sinc}(\frac{\pi x}{h})=\delta_h(x),$$
where $\mathcal{F}$ denotes the Fourier transform and $\mathcal{F}^{-1}$ denotes the inverse Fourier transform. Thus, it turns out the inverse Fourier transform of $\tilde{\delta}_h(k)$ is the `$\mathtt{sinc}$' kernel $\delta_h(x)$ and this is exactly why it is denoted as $\tilde{\delta}_h$ as the $\tilde{\cdot}$ represents the Fourier transform operator. Note that in the previous context we use $\hat{\cdot}$ to represent Fourier series coefficients. Therefore, the convolutions of the `$\mathtt{sinc}$' kernel should be an integral in the whole space as

\begin{align*}
&f(x)=\sum_{n=-\infty}^{\infty}\int_{\theta \in \Gamma}F(\theta)\delta_{h}(x-(X(\theta)+nL))d \theta,\\
&U( \theta ) = \int_{-\infty}^{\infty} u(x, t) \delta_h(x-X(\theta)) d x .\\
\end{align*}

Note that the `$\mathtt{sinc}$' kerne $\delta_h$ is a function  that is non-periodic and defined in the whole space and the $\delta_{L,N}$ kernel is the `periodized' version of $\delta_h$. The difference before and after periodization in the standard IB method is minor because the IB kernel $\delta_h^{(n)}$ is supported only on several meshwidth. But it is not the case for the global FSIB kernel and it is important to specify whether it is an integral over the whole space or only one period.  

It is easy to show that the `$\mathtt{sinc}$' kernel $\delta_h$ satisfies the IB kernel moment conditions (\romannumeral 1) and (\romannumeral 3) in \cref{eq:tradIBkernels} in the form of continuous integral, i.e. $\int_{\mathbb{R}} \delta_h(x) \,dx=1$ and $\int_{\mathbb{R}} x^{\alpha}\delta_h(x) \,dx=0$ where $\alpha=1,2,\ldots$. This is because any $\alpha$ order moment in physical space is the $\alpha$ order derivative evaluated at $k=0$ in Fourier space ,and we know the fact that $\tilde{\delta}_h(0)=1$ and $\tilde{\delta}^{(\alpha)}_h(0)=0$. 

Note that the `$\mathtt{sinc}$' kernel satisfies the discrete IB kernel conditions in \cref{eq:tradIBkernels} as well as the corresponding continuous conditions. Consider any function $f: \mathbb{R}\rightarrow \mathbb{R}$. The function is periodized with any period $a$ as $\sum_{j=-\infty}^{\infty}f(x-ja)$. Since it is periodic and thus we can find its Fourier coefficients as

\begin{equation}
\sum_{j=-\infty}^{\infty} f(x-j a)=\sum_{k=-\infty}^{\infty} c_k e^{2 \pi i k \frac{x}{a}},    
\end{equation}
where 
\begin{equation}
\begin{aligned}
c_k&=\frac{1}{a} \int_0^{a} \sum_{j=-\infty}^{\infty} f(x-j a) e^{-2 \pi i k \frac{x}{a}} d x \\
&=\frac{1}{a} \sum_{j=-\infty}^{\infty} \int_0^a f(x-j a) e^{-2 \pi i k \frac{x-j a}{a}} d x \\
&=\frac{1}{a} \int_{-\infty}^{\infty} f(x) e^{-2 \pi i k \frac{x}{a}} d x=\frac{1}{a} \tilde{f}\left(\frac{2 \pi k}{a}\right).
\end{aligned}
\end{equation}
Thus, we know that the periodized function can be written as

\begin{equation}\label{eq:peri_four_coef}
\sum_{j=-\infty}^{\infty} f(x-j a)=\frac{1}{a} \sum_{k=-\infty}^{\infty} \tilde{f}\left(\frac{2 \pi k}{a}\right) e^{2 \pi i k \frac{x}{a}} .
\end{equation}
This identity helps us derive interesting properties of $\delta_h$. Most obviously, let $f=\delta_h$ and $a=L$, then 

\begin{equation}
\sum_{j=-\infty}^{\infty} \delta_h(x-j L)=\frac{1}{L} \sum_{k=-\infty}^{\infty} \tilde{\delta}_h\left(\frac{2 \pi k}{L}\right) e^{2 \pi i k \frac{x}{L}}=\frac{1}{L}\sum_{k=-\frac{N-1}{2}}^{\frac{N-1}{2}}e^{2\pi i k \frac{x}{L}}=\delta_{L,N}(x),
\end{equation}
where the second identity comes from the fact that $\tilde{\delta}_h$ is a square function. Recall that $\delta_h$ can be obtained by the limit of $\delta_{L,N}$ as $L,N\rightarrow \infty$ with $L/N=h$. Here we prove that $\delta_{L,N}$ can be reconstructed from $\delta_h$ by periodization, which we stated before without proof.

Another application of \cref{eq:peri_four_coef} is to prove the moment conditions. Let $a=h$ and let $f(x)=x^\alpha \delta_h(x)$ for any non-negative integer $\alpha$, then

\begin{equation}
\sum_{j=-\infty}^{\infty}(x-j h)^\alpha \delta_h(x-j h) h=\sum_{k=-\infty}^{\infty}\widetilde{\left(x^\alpha \delta_h\right)}\left(\frac{2 \pi k}{h}\right) e^{2 \pi i k \frac{x}{h}}=\left\{
\begin{aligned}
&1,\quad \alpha=0\\
&0,\quad \alpha=1,2,3,...
\end{aligned}
\right.
\end{equation}
When $\alpha=0$, the term in the summation is $1$ only when $k=0$ and is $0$ otherwise, so we prove the zeroth moment condition (\romannumeral 1). When $\alpha\ge 1$, it is always zero because the $\alpha$ order moment is proportional to the $\alpha$ order derivative of $\tilde{\delta}_h$ which is zero. So we prove that the moment conditions in \cref{eq:tradIBkernels} are all satisfied for the `$\mathtt{sinc}$' kernel $\delta_h$. 

\begin{figure}
    \centering
    \includegraphics[width=0.7\textwidth]{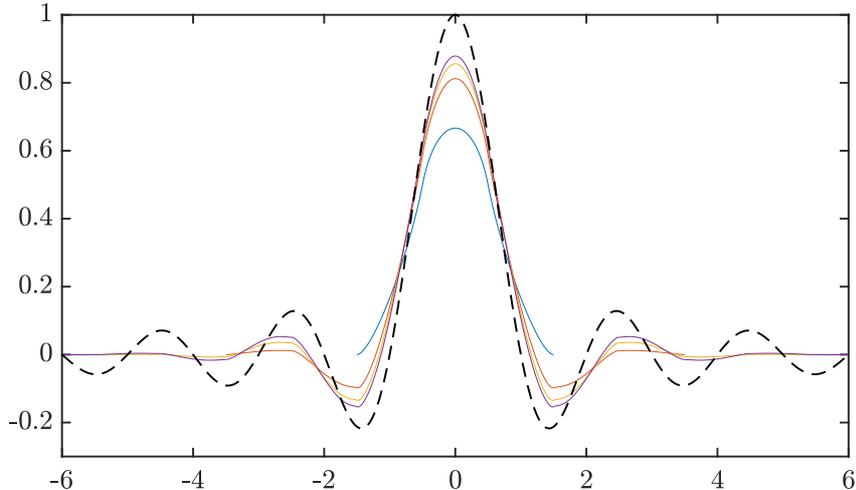}
    \caption{ $\delta_h(x)\cdot h$ vs $x/h$. The new `$\mathtt{sinc}$‘ kernel (the black dashed line) is compared with standard kernels (the solid lines) with an odd number of points. 3-point kernel (blue), 7-point kernel (red), 11-point kernel (yellow), and 15-point kernel (purple).}
    \label{fig:kernels_comp}
\end{figure}

So we have checked all the IB conditions for odd $n$, i.e. the condition (\romannumeral 1), (\romannumeral 3), and (\romannumeral 4) are satisfied by the FSIB kernel. Recall that for every $n$, these conditions uniquely determined an IB kernel $\delta_{h}^{(n)}$. Note that the IB kernels are finitely supported but the FSIB kernel is global. Thus it is reasonable to make a conjecture that  

\begin{equation}
\lim _{\substack{n \rightarrow \infty \\ n \text { odd }}} \delta_h^{(n)}(x)=\delta_h(x).
\end{equation}
We give a visual demonstration of this in \cref{fig:kernels_comp} where the standard IB kernels with $n=3,7,11,15$ along with the FSIB `$\mathtt{sinc}$' kernel are shown. Note that we make a nondimensionalization by plotting $\delta_h(x)\cdot h$ vs $x/h$. As $n$ gets bigger, the standard IB kernel develops more wiggles, has wider support, and is closer to the FSIB `$\mathtt{sinc}$' kernel. Also, the central peak is narrower and narrower and it represents the effective width of the immersed boundary. Note that the central peak of the `$\mathtt{sinc}$' kernel almost has a width of only one meshwidth $h=L/N$, which is the best we can achieve numerically. This is how it improves singularity resolution on the immersed boundary.

\vspace{6pt}

\textbf{Remark 1.} One may question that the quadrature makes \cref{eq:force_spreading_quad} equivalent to compute Fourier transform of $N_b$ discrete singular force points, i.e. $\sum_{q=1}^{N_b} \omega_q \bs F(\bs \theta_q)\delta (\bs x-\bs X(\bs \theta_q))$. It seems not only to make the force singular in the direction perpendicular to the immersed boundary, which is what we desire, but also to make it singular along the immersed boundary, which is not physical. For example, a singular point force will generate infinite flow near the singularity in Stokes flow. For Navier-stokes flow, singular point force on the immersed boundary will cause the flow to leak between the singular points. As discussed above, the new method replaces the regularized delta functions used in the IB method with the $\mathtt{sinc}$ function. Therefore, it smooths the singular boundary. More importantly, the effective width of the singular immersed boundary in the FSIB method is equal to $L/N$ approximately, which is the best we can do numerically. This provides some insight into how to choose the number of Fourier modes $N$ and the number of Lagrangian grid points $N_b$ accordingly. The number of Fourier modes determines the resolution in physical space as $L/N$ and it should match the Lagrangian meshwidth $h_b$. If $h_b\ll L/N$, the Fourier space is not well-resolved and still has a large aliasing error. As we add Fourier modes, the aliasing error decreases. Nevertheless, it should not cross the match point because the regularized point force has an effective radius of about $L/N$ and the immersed boundary will not be continuous and causes severe leakage if $h_b\gg L/N$. This will be shown in our numerical examples in \cref{sec:numerical}. 

\section{Properties of the FSIB method} \label{sec:prop_FSIBM}

\subsection{Divergence-free condition and volume conservation.}

An advantage of the FSIB method is that the velocity field is analytically divergence-free. Enforced by the Fourier spectral method, we automatically have $\hat{\bs u}(\bs k)\cdot \bs k=0$ in Fourier space. The velocity field $\bs U(\bs x)$ at any point $\bs x$ is interpolated by evaluation of Fourier series in  \cref{eq:force_spreading_quad}. Thus, we know that $\nabla \cdot \bs U=0$ analytically everywhere in physical space.

As discussed in the introduction, Bao et al. \citep{baoImmersedBoundaryMethod2017} have proposed an IB method with a divergence-free velocity field. In that method, a vector potential is introduced and the velocity field $\bs U$ is computed by the curl of the vector potential, thus satisfying the analytical divergence-free condition. Our new Fourier spectral method achieves the same result without the use of a vector potential. A direct consequence of the divergence-free velocity is volume conservation. We observe that the volume leakage is eliminated in numerical tests in \cref{sec:numerical}.

\subsection{Translation invariance} \label{sec:trans_invar}

Exact translation invariance cannot be achieved in the standard IB method because the regularized delta function has finite support. Here, however, we have a kernel with unbounded support, see \cref{sec:relation}. We will prove the translation invariance for Stokes flow and for Navier-Stokes flow separately in the following sections.

\subsubsection{Exact translation invariance for Stokes flow}

Consider a force density $\bs f$ that is moved by displacement $\bs r$, then denote the new force density $\bs f_r(\bs x)=\bs f(\bs x-\bs r)$. The exact translation invariance means that the solution of our new method, denoted as $\bs u_r$, should also move the same displacement, i.e. $\bs u_r(\bs x)=\bs u(\bs x-\bs r)$.

Denote the truncation operator in Fourier space as $\mathcal{T}_{N}$. In Fourier space, it is convenient that the displacement simply becomes a phase shift as 

\begin{equation*}
\hat{\bs f}_r(\bs k)=\hat{\bs f}(\bs k) \exp(-i\bs k \cdot \bs r).
\end{equation*}

Also, the spectral solver in Fourier space simply becomes the Stokeslet kernel in \cref{eq:stokeslet}, denoted as operator $\mathcal{M}$. Thus, we get the solution as 

\begin{equation*}
\hat{\bs u}_r(\bs k) = \mathcal{T}_{N} \circ \mathcal{M} \circ \mathcal{T}_{N} \circ \hat{\bs f}_r (\bs k).
\end{equation*}

The mobility operator $\mathcal{M}$ is diagonalized in Fourier space and thus we have $\hat{\bs u}_r(\bs k)=\hat{\bs u}(\bs k) \exp(-i\bs k \cdot \bs r)$. This proves the desired exact translational invariance $\bs u_r(\bs x)=\bs u(\bs x-\bs r)$. 

Note that the duality of the spreading and interpolation operator is shown vividly above as they are the same truncation operator $\mathcal{T}_{N}$. The last truncation operator, however, does nothing in the case of Stokes flow because the output of $\mathcal{M}$ operator has the same Fourier modes as the input. But it is not the case when it is Navier-Stokes flow.

\subsubsection{Exact translation invariance for Navier-Stokes flow}

For Navier-Stokes flow, at each time step, we are given $\hat{\bs u}$ and $\hat{\bs f}$. So, the translation gives $\bs u_r(\bs x,t)=\bs u(\bs x-\bs r,t)$ and $\bs f_r(\bs x)=\bs f(\bs x-\bs r,t)$. Similarly, we have the phase shifted function in Fourier space as $\hat{\bs u}_r(\bs k,t)=\hat{\bs u}(\bs k,t) \exp(-i\bs k \cdot \bs r)$ and $\hat{\bs f}_r(\bs k,t)=\hat{\bs f}(\bs k,t) \exp(-i\bs k \cdot \bs r)$. We need to prove that the solution of the FSIB method, i.e. the time derivative $\partial \bs u/\partial t$, at any location satisfies the translation condition that 
\begin{equation}\label{eq:trans_cond}
\frac{\partial \bs u_r(\bs x,t)}{\partial t}=\frac{\partial \bs u (\bs x-\bs r,t)}{\partial t}.
\end{equation}

We denote the spectral solver in \cref{eq:fourier_spectral_sol} as an operator $\mathcal{V}$. Thus, 

\begin{equation*}
\frac{\partial \hat{\bs u}_r(\bs x,t)}{\partial t} =  \mathcal{T}_{N} \circ \mathcal{V} \circ \mathcal{T}_{N} \circ \hat{\bs f}_r (\bs x).
\end{equation*}

Note that all the linear terms in the spectral solver preserve the factor $\exp(-i\bs k \cdot \bs r)$ so do not compromise the identity \cref{eq:trans_cond}. To prove the translation condition in \cref{eq:trans_cond}, we only need to prove the nonlinear term $\hat{s}_r(\bs k) = \hat{s}(\bs k) \exp (-i\bs k \cdot \bs r)$. By the convolution theorem, the multiplication in physical space becomes convolution in Fourier space such that 

\begin{equation*}
\begin{array}{ll}
\hat{s}_r(\bs k) &= \sum_{\bs k_{1}} \hat{\bs u}_r (\bs k- \bs k_1) \cdot i k_{1} \hat{\bs u}_{r}\left(\bs k_{1}\right)\\
&=\sum_{\bs k_{1}} \hat{\bs u}(\bs k-\bs k_{1}) \cdot i \bs k_{1} \hat{\bs u} (\bs k_{1}) \cdot \exp \left(-i(\bs k-\bs k_{1} )\cdot \bs r \right)\cdot \exp \left(-i \bs k_{1} \cdot \bs r \right) \\
&=\sum_{\bs k_{1}} \hat{\bs u}\left(\bs k-\bs k_{1}\right) \cdot i \bs k_1 \hat{\bs u}\left(\bs k_{1}\right) \exp (-i \bs k \cdot \bs r)\\
&=\hat{s}(\bs k) \cdot \exp (-i \bs k \cdot \bs r). \\
\end{array}
\end{equation*}
Thus, we prove the translation invariant condition in \cref{eq:trans_cond}. 

The duality of the spreading and the interpolation operators is also satisfied. Moreover, the last truncation in Fourier space takes effect because the nonlinear term will create higher frequency modes and we leave them out by the last truncation operator. This is different from the Stokes case where it has no effect.

The above proof of translation invariance is in the setting of continuous time. Thus, the exact translation invariance for an arbitrary temporal integrator is not guaranteed, but at least the translation invariance holds in the limit $\Delta t \rightarrow 0$.

\subsection{Conservation of momentum}

Note that the integral of a function in physical space is the zero frequency in Fourier space, which gives

\begin{equation}
\begin{aligned}
&\hat{\bs f}(0, t)=\frac{1}{L^{3}} \iiint_{\Omega_L} \bs f(\bs x,t) d\, \bs x = \frac{1}{L^{3}} \sum_{j=1}^{N_b} \omega_j \bs F_{j} (t),\\
&\hat{\bs u}(0, t)=\frac{1}{L^{3}} \iiint_{\Omega_L} \bs u(\bs x,t) d\, \bs x .
\end{aligned}
\end{equation}
Therefore, to prove the conservation of momentum, we just need to find the equation of the zero frequency of the velocity and the force in Fourier space. Setting $\bs k=0$ in \cref{eq:fourier_spectral_sol} and we immediately obtain the equation for zero frequency.

\begin{equation}
\rho\left(\frac{\partial \hat{\bs u}}{\partial t}(0, t)+\hat{\bs s}(0, t)\right)=\hat{\bs f}(0, t).
\end{equation}
Note that the term of $\bs k \bs k^T/(|\bs k|^2)$ is eliminated because we set it to be zero when $\bs k=0$. Then, it is easy to prove the nonlinear term is zero $\hat{\bs s}(0,t)=0$.

\begin{equation}
\begin{aligned}
\hat{\bs s}_{\alpha}(0, t) &= \frac{1}{L^{3}} \iiint_{\Omega_L} \bs u (\bs x, t) \cdot\left(\nabla \bs u_{\alpha}\right)(\bs x, t) d \bs x \\
&=\frac{1}{L^{3}} \iiint_{\Omega_L} \nabla \cdot\left(\bs u(\bs x, t) \bs u_{\alpha}(\bs x, t)\right) d \bs x=0,
\end{aligned}
\end{equation}
where $\alpha = 1,2,3$ and we have used the divergence-free condition $\nabla \cdot \bs u = 0$ to integrate by part, and periodicity to conclude the last integral is zero.

Now, all ingredients are prepared to prove the conservation of momentum, which is to prove the statement that the rate of change in time of fluid momentum is equal to the total force applied to the fluid.

\begin{equation}
\frac{d}{d t} \left( \iiint_{\Omega_L} \rho \bs u (\bs x,t) d \bs x \right) = \frac{d}{d t} \left( \rho L^3 \hat{\bs u} (0,t) \right) = \sum_{j=1}^{N_b} \omega_j \bs F_{j} (t)
\end{equation}

\subsection{Conservation of energy}

First, the kinetic energy of the fluid in each period is given by

\begin{equation}
\begin{aligned}
E_K(t) &=\frac{1}{2} \rho \iiint_{\Omega_L} \bs u (\bs x, t) \cdot \bs u(\bs x, t) d \bs x \\
&=\frac{1}{2} \rho L^3 \sum_{\bs k \in \mathcal{K}_N} \hat{\bs u}(\bs k, t) \cdot \bs \hat{u}(-\bs k, t).
\end{aligned}
\end{equation}
So we know the rate of change of the kinetic energy is  

\begin{equation}
\begin{aligned}
\frac{d E_K}{d t}(t)=& \frac{1}{2} \rho L^3 \sum_{\bs k \in \mathcal{K}_N}\left(\frac{\partial \hat{\bs u}}{\partial t}(\bs{k}, t) \cdot \hat{\bs u}(-\bs{k}, t)+\hat{\bs{u}}(\bs{k}, t) \cdot \frac{\partial \hat{\bs u}}{\partial t}(-\bs{k}, t)\right) \\
=& \rho L^3 \sum_{\bs k \in \mathcal{K}_N} \frac{\partial \hat{\bs u}}{\partial t}(\bs{k}, t) \cdot \hat{\bs u}(-\bs{k}, t).
\end{aligned}
\end{equation}
From the moment condition of the Navier-Stokes \cref{eq:navierstokes} in Fourier space, we know 

\begin{equation}\label{eq:moment_k_space}
\rho\left(\frac{\partial \hat{\bs{u}}}{\partial t}(\bs{k}, t) \cdot \hat{\bs{u}}(-\bs{k}, t)+\hat{\bs{s}}(\bs k, t) \cdot \hat{\bs{u}}(-\bs{k}, t)\right)=-\mu\|\bs{k}\|^2 \hat{\bs{u}}(\bs{k}, t) \cdot \hat{\bs{u}}(-\bs{k}, t)+\hat{\bs{f}}(\bs{k}, t) \cdot \hat{\bs{u}}(-\bs{k}, t),
\end{equation}
where we multiply both side by $\hat{\bs u}(-\bs k,t)$. If we sum this over $\bs k \in \mathcal{K}_N$, then the first term on the LHS is equal to $d E_k/d t$ and we will prove later that the second term involving the nonlinear term vanishes. Therefore, we have

\begin{equation}\label{eq:kine_energ_derivative}
\frac{d E_K}{d t}(t)=-\mu L^3 \sum_{\bs{k} \in \mathcal{K}_N}\|\bs{k}\|^2 \hat{\bs{u}}(\bs{k}, t) \cdot \hat{\bs{u}}(-\bs{k}, t)+L^3 \sum_{\bs{k} \in \mathcal{K}_N} \hat{\bs f}(\bs{k}, t) \cdot \hat{\bs{u}}(-\bs{k}, t).
\end{equation}
Note that the first term on the RHS is the rate of viscous energy dissipation. So, in order to prove the conservation of energy, we only need to prove the second term on the RHS is the negative rate of change of the potential energy $E_p$. By the force spreading (\cref{eq:force_spreading_quad}) and the velocity interpolation (\cref{eq:velo_intp_fini}), we have

\begin{equation}\label{eq:pote_energ_derivative}
\begin{aligned}
L^3 \sum_{\bs{k} \in \mathcal{K}_N} \hat{\bs{f}}(\bs{k}, t) \cdot \bs{\hat{u}}(-\bs{k}, t)&=\sum_{j=1}^{N_b} \omega_j \bs F_{j}(t) \cdot \sum_{\bs{k} \in \mathcal{K}_N} e^{-i \frac{2 \pi}{L} \bs{k} \cdot \bs X_{j}(t)} \hat{\bs{u}}(-\bs{k}, t)\\
&=\sum_{j=1}^{N_b} \omega_j \bs F_{j} (t) \sum_{\bs{k} \in \mathcal{K}_N} e^{i \frac{2 \pi}{L} \bs k \cdot \bs X_{j}(t)} \hat{\bs u}(\bs{k}, t)\\
&=\sum_{j=1}^{N_b} \omega_j \bs F_{j}(t) \cdot \bs U_j(t)\\
&=\sum_{j=1}^{N_b} -\frac{\partial E_p}{\partial \bs X_j} \cdot \frac{\partial X_j}{\partial t}\\
&=-\frac{d E_p}{dt}.
\end{aligned}
\end{equation}
Note that we make the change from $\hat{\bs u}(-\bs k,t)$ to $\hat{\bs u}(\bs k,t)$ because $\hat{\bs u}(\bs k,t)$ is even in $\bs k$ for a real function $\bs u$. We also use the fact that the force on the immersed boundary is given by the negative derivative of the potential, i.e. $\omega_j \bs F_{j}=-\partial E_p/\partial \bs X_j$. By substituting \cref{eq:pote_energ_derivative} into \cref{eq:kine_energ_derivative}, we get the energy conservation for the spatially discretized system in the FSIB method as

\begin{equation}
\frac{d}{d t}\left(E_K+E_p\right)=-\mu \sum_{\bs{k} \in \mathcal{K}_N}\|\bs{k}\|^2 \bs{\hat{u}}(\bs{k}, t) \cdot \bs{\hat{u}}(-\bs{k}, t).
\end{equation}
There is one last thing we need to show to complete the proof though. We need to prove the nonlinear term in \cref{eq:moment_k_space} will vanish after summing up over $\bs k \in \mathcal{K}_N$. So we have

\begin{equation}
\begin{aligned}
\sum_{\bs{k} \in \mathcal{K}_N} \bs{\hat{s}} (\bs{k}, t) \cdot \bs{\hat{u}}(-\bs{k}, t) & =\sum_{\bs{k} \in \mathcal{K}} \hat{\bs{s}} (\bs{k}, t) \cdot \bs{u}(-\bs{k}, t)\\
&=\frac{1}{L^3} \iiint_{\Omega_L} \bs{u} \cdot(\bs{u} \cdot \nabla \bs{u}) d \bs{x}\\
&=\frac{1}{L^3} \iiint_{\Omega_L} \sum_{\alpha=1}^3 u_\alpha \bs{u} \cdot \nabla u_\alpha d \bs{x}\\
&=\frac{1}{L^3}\sum_{\alpha=1}^3 \iiint_{\Omega_L}   u_\alpha  \nabla \cdot(\bs u u_\alpha ) d \bs{x}\\
&=-\frac{1}{L^3}\sum_{\alpha=1}^3 \iiint_{\Omega_L}  \nabla u_\alpha  \cdot u_\alpha \bs{u} d \bs{x}\\
&=0,
\end{aligned}
\end{equation}
where we extend $\mathcal{K}_N$ to all the Fourier modes $\mathcal{K}$ in the first line because $\hat{\bs s}=0$ outside of $\mathcal{K}_N$. We also use the fact that $\nabla \cdot \bs u=0$ in the fourth line. An integral by part is used in the fifth line. The identity is zero because the third line and the fifth line are opposite to each other so the value must be zero. Thus we now complete the proof of the conservation of energy.

\section{Numerical results}\label{sec:numerical}

\subsection{Stokes equations}\label{sec:stokes}

A good starting point for our numerical experiments is the incompressible flow in two dimensions governed by the Stokes equations

\begin{equation}\label{eq:stokes}
\left\{\begin{array}{l}
-\nabla p+\mu \Delta \boldsymbol{u}+\boldsymbol{f}=0, \\
\nabla \cdot \boldsymbol{u}=0.
\end{array}\right.
\end{equation}

In this case, there is no nonlinear term or time integrator. For comparison with the standard IB method, we are interested in the FSIB method's performance and properties for a benchmark model that a closed elastic neutral-buoyant curve $\Gamma =  \{\bs x: \bs x=\bs X(\theta)\}$, $\theta=[0,2\pi)$ is immersed in the flow. Similarly, we set it periodic in both directions as a $[0,L]\times[0,L]$ periodic box.  The force $F(\theta)$, $\theta=[0,2\pi)$ is given on the closed boundary $\Gamma$ as a singular layer. Thus, the force vector field can be written as integral with the Dirac delta function

\begin{equation}\label{eq:force}
    f(x)=\int_{0}^{2 \pi} F(\theta) \delta(x-X(\theta)) d \theta.
\end{equation}

Moreover, for the existence of the Stokes solution, it is required that the integral of force inside each periodic box vanishes.

\begin{equation}\label{eq:force_vanish}
\int_0^{2\pi}F(\theta)d\theta =0
\end{equation}

The Stokes equations (\ref{eq:stokes}) in a periodic box are easily solved by Fourier series.  The solution in Fourier space is

\begin{equation}\label{eq:stokeslet}
\left\{\begin{array}{l}
\hat{u}=\frac{1}{\mu}\left(\frac{1}{|k|^{2}}-\frac{k k^T}{|k|^{4}}\right) \hat{f} \\
\hat{p}=\frac{k \cdot \hat{f}}{i|k|^{2}}
\end{array}\right. ,
\end{equation}
which is the term-by-term product of the Fourier series of the Stokeslet and the Fourier series of the applied force. This follows from the linearity of the Stokes equations and the convolution theorem for Fourier series.

\begin{figure}[!htbp]
    \centering
    \begin{subfigure}[b]{0.35\textwidth}
        \centering
        \includegraphics[width=\textwidth]{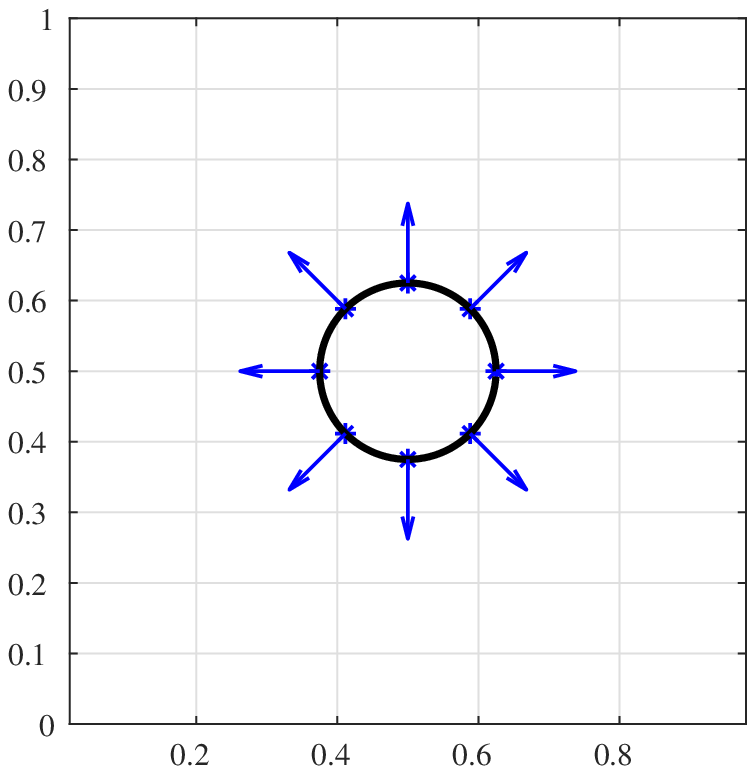}
        \caption{Normal case.}\label{fig:circle:normal}
    \end{subfigure}
    \begin{subfigure}[b]{0.35\textwidth}  
        \centering 
        \includegraphics[width=\textwidth]{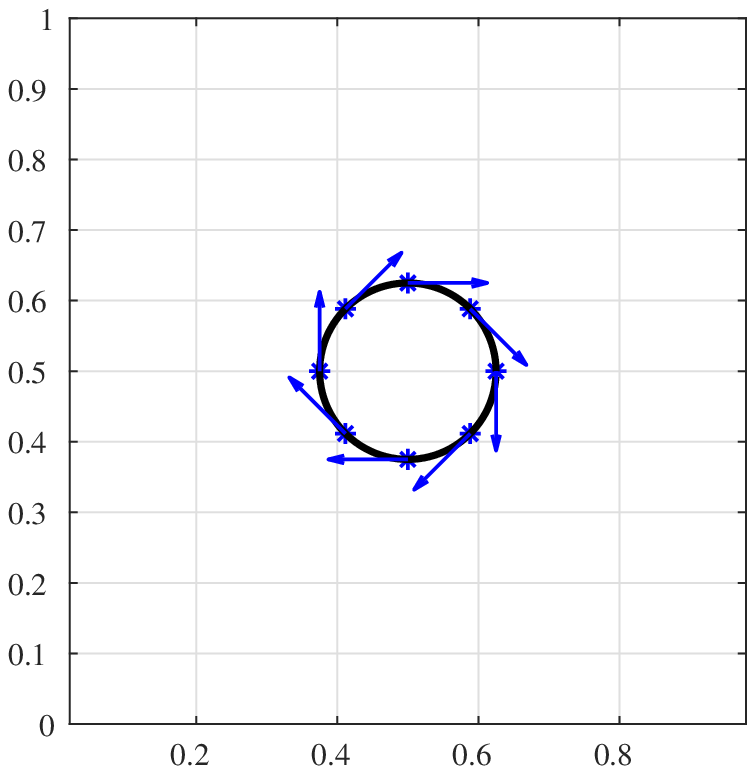}
        \caption{Tangent case.}\label{fig:circle:tangent}
    \end{subfigure}
    \caption{Circle problem for Stokes flow. For (a) `normal' case, a uniform force field is applied to the circle in the radial outwards direction. For (b) `tangent' case, a uniform force field is applied tangent to the circle clockwise.}\label{fig:circle}
\end{figure}

We first consider an immersed boundary in the form of a circle (\cref{fig:circle}). The circle has a radius of $r=L/8$ and is located at the center of the square box with a length $L=1$. The viscosity coefficient is set to be $\mu=1$. 
In the first `normal' case (\cref{fig:circle:normal}), there is an outward normal force, applied along the circular immersed boundary, and this results in a pressure difference across the boundary. In the second `tangent' case (\cref{fig:circle:tangent}), the applied force is tangent to the immersed boundary. In both cases, the magnitude of the force per unit $\theta$ is set to be constant $|\bs F|=1$. Thus, the net-zero force condition in \cref{eq:force_vanish} is automatically satisfied and solutions to the Stokes equations therefore exist. Our discretization of the Stokes equations is derived from the one used for the Navier-Stokes equations in \cref{sec:discretization} simply by removing the nonlinear and also the time-derivative terms. Since the fluid domain is now two-dimensional, the Lagrangian parameter $\theta$ is now one-dimensional and can be discretized as $\theta_j=j\cdot \Delta \theta$, where $j=0,...,N_b-1$ and $\Delta \theta=2\pi /N_b$. Therefore, the Lagrangian markers have coordinates $\bs X_j=(r\cos(\theta_j),r\sin(\theta_j))$.

For the `normal' case, the analytical solution of velocity should be zero everywhere despite the lack of radial symmetry on our periodic domain. The reason for this is that the singular force field considered here is a gradient (of a step function, with one value inside and another value outside of the circular immersed boundary), and therefore its Hodge projection onto the space of incompressible vector fields is zero, which we also verify from the numerical tests. There will be a pressure jump on the circle since there is normal force uniformly distributed. The pressure is constant inside and outside of the circle because the force vanishes there. Now assume that the integral of the pressure over the periodic box is zero, which we may do because addition of a constant to the pressure makes no difference. The exact solution for the pressure is then given by

\begin{equation*}
\left\{
\begin{array}{l}
p_1=\frac{F}{r}\frac{\pi r^2}{L^2},\\
p_2=-\frac{F}{r}\frac{L^2-\pi r^2}{L^2},
\end{array}
\right. 
\end{equation*}
where $p_1$ is the pressure outside of the circle, and $p_2$ is the pressure inside of the circle.

\begin{figure}[!htbp]
  \centering
  \begin{subfigure}[b]{0.45\textwidth}
  \includegraphics[width=\textwidth]{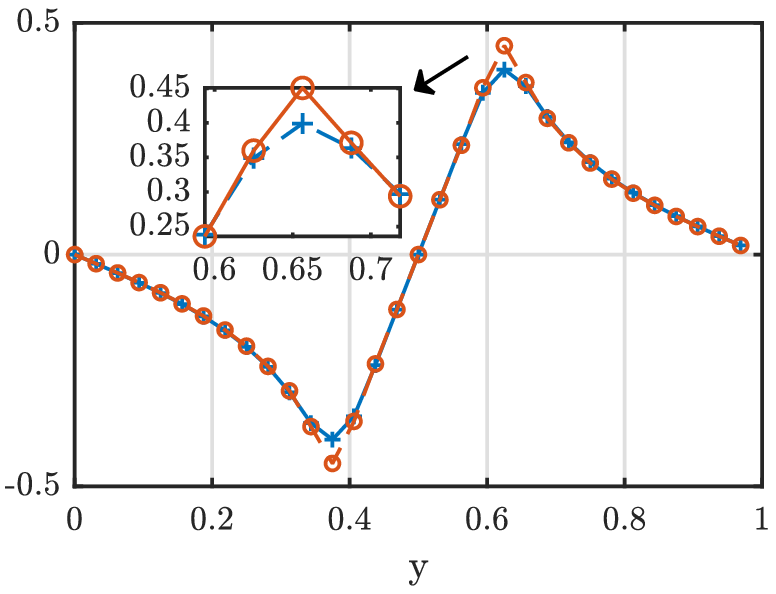}
    \caption{}\label{fig:circle:u2_centerline}
  \end{subfigure}
  \hfill
  \begin{subfigure}[b]{0.45\textwidth}
  \includegraphics[width=\textwidth]{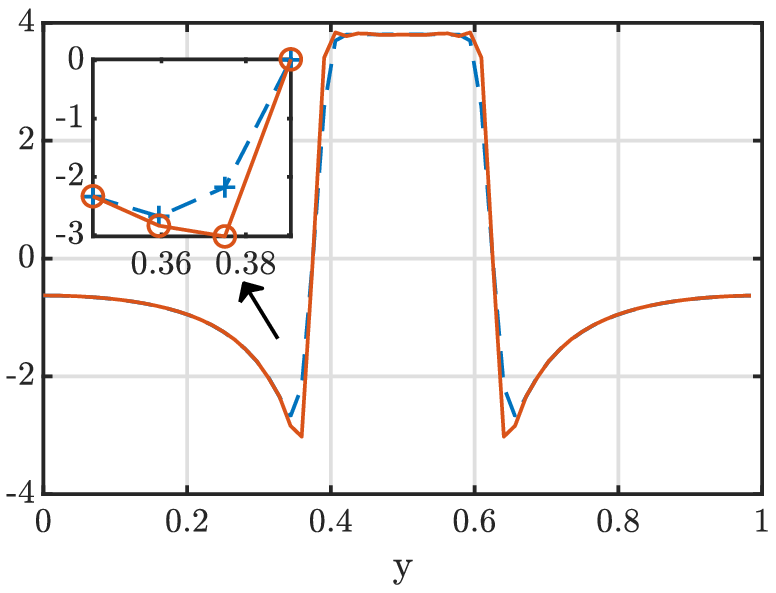}
    \caption{}\label{fig:circle:u2_p_centerline}
  \end{subfigure}
  \caption{(a). Velocity $u_x$. (b). Derivative of the velocity $d u_x/dy$. Both are plotted on the vertical centerline, i.e $x=L/2$, as a function of $y$ for the `tangent' case. The red solid lines with markers `o' are for the FSIB method. The blue dashed lines with markers `+' are for the standard IB method.}
\end{figure}

For the `tangent' case, the pressure should be constant, which is not a trivial result. The reason is that $\nabla \cdot \bs f=0$. A complete analysis of the jump condition is given in \citep{laiRemarkJumpConditions2001}. There will be a circular flow around the circle inside or outside. We plot the numerical result of velocity $u_x$ on the vertical centerline, i.e. $x=L/2$, as a function of $y$ in \cref{fig:circle:u2_centerline}. We also plot the numerical result of the standard IB method with the 4-pt standard kernel for the same grids, i.e. same $N_b$ and $N$, for comparison, but here $N$ is the number of Eulerian grid points rather than the number of Fourier modes in the FSIB method. Remarkably, the sharpness of the boundary for the new FSIB method is visibly better than the standard IB method. It is no surprise that the flow inside of the circle is almost the same as a rigid body rotation, which means $u_x$ is linear in $y$ on the vertical centerline. But it is not an exactly rigid body rotation since the periodic boundary condition breaks the rotational symmetry and this has a small effect inside of the immersed boundary as well as on the outside. Moreover, we see no oscillation on the spikes since there is no jump discontinuity of velocity. Although the analytical solution of the velocity is not available, we know that the derivative of velocity $du_x/dy$ on the vertical centerline should have a jump discontinuity over the circular boundary as a result of the uniform tangential force. We want to observe how the new FSIB method behaves for the jump discontinuity, so we plot $du_x/dy$ on the vertical centerline for both the new FSIB method and the standard IB method for the same grids in \cref{fig:circle:u2_centerline}. This figure shows that the new FSIB method has a better resolution of the jump discontinuity. More quantitative comparisons of the boundary resolution will be made in \cref{sec:bdr_reso_stokes} for Stokes flow and in \cref{sec:poiseuille} for Navier-Stokes flow.

\begin{figure}[!htbp]
    \centering
    \includegraphics[width=.7\textwidth]{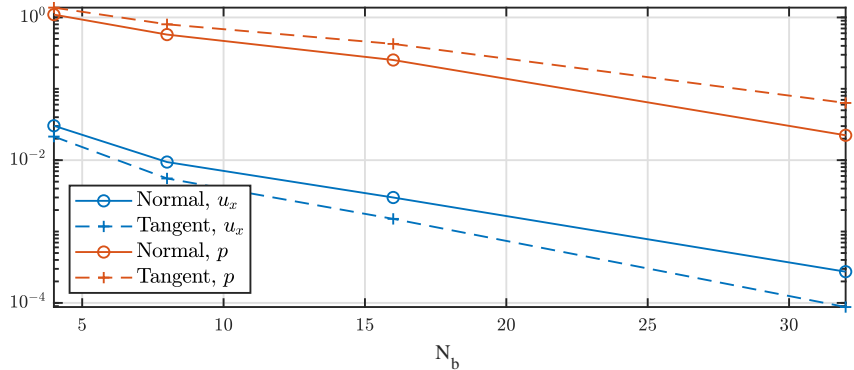}
    \caption{Log-linear scale plot. Error in norm 2 of numerical solutions of pressure $p(y)$ compared with analytical solutions and empirical error in norm 2 of velocity $u_x(y)$ on the vertical centerline, i.e. $x=L/2$. For fixed $N=128$, we increase $N_b=4,8,16,32$. Blue lines are for velocity $u_x$, red lines are for pressure $p$. Solid lines with markers `o' are for the `normal' case and dashed lines with markers `+' are for the `tangent' case.}
    \label{fig:err_2_N}
\end{figure}

As is discussed in \cref{sec:spatial_discretization}, the spreading of the force to the uniform grids (\cref{eq:force_spreading_quad}) is exponentially accurate for a simple trapezoidal rule because the function $F$ is smooth as a function of the Lagrangian variable $\theta$. Here we design a numerical experiment to verify such exponential accuracy. We fix the number of Fourier modes $N=128$ and increase the number of points on the Lagrangian grids $N_b=4,8,16,32$. In \cref{fig:err_2_N}, we plot the error of velocity $u_N$ and pressure $p_N$ on the vertical centerline as a function of $N_b$. The errors of both the pressure and the velocity converge exponentially for the `normal' case and also for the `tangent' case.

\subsubsection{Coupling of the refinement of the Fourier resolution and the Lagrangian meshwidth}

\begin{figure}[!htbp]
  \centering
  \begin{subfigure}[b]{0.32\textwidth}
  \includegraphics[width=\textwidth]{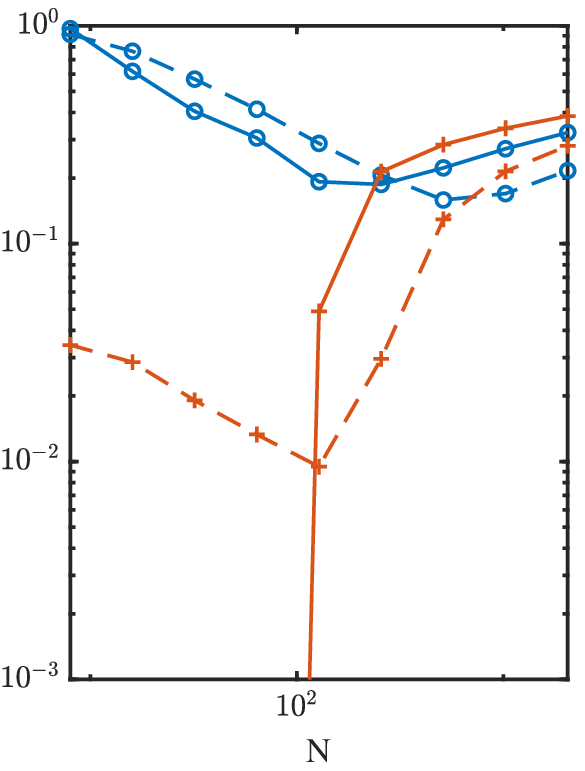}
    \caption{}\label{fig:circle:p_Nh_comp}
  \end{subfigure}
  \hfill
  \begin{subfigure}[b]{0.32\textwidth}
  \includegraphics[width=\textwidth]{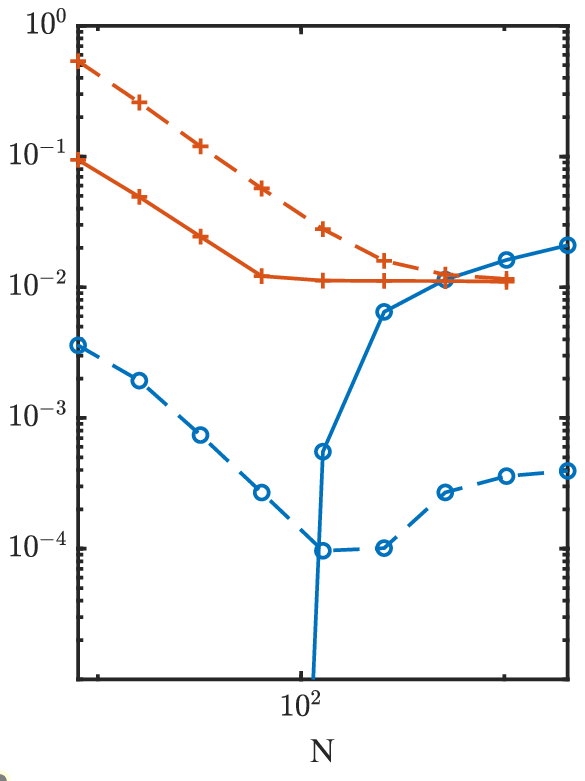}
    \caption{}\label{fig:circle:u_Nh_comp_bd}
  \end{subfigure}
  \hfill
  \begin{subfigure}[b]{0.32\textwidth}
  \includegraphics[width=\textwidth]{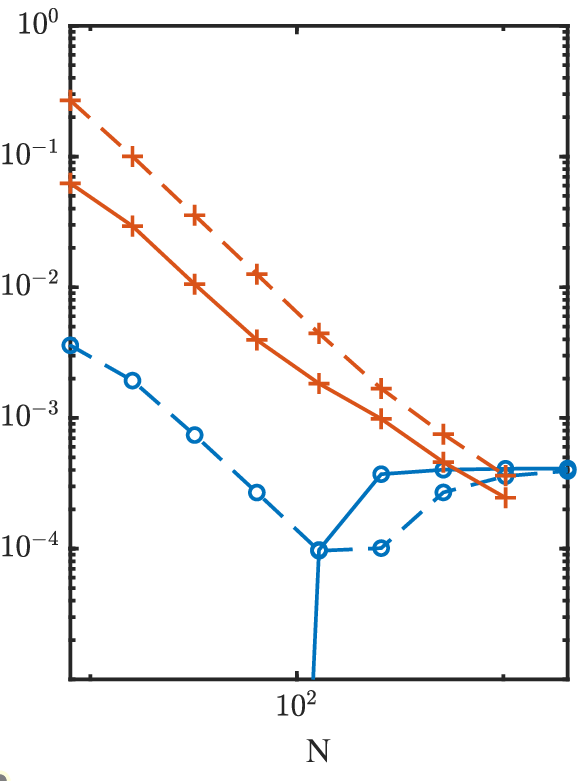}
    \caption{}\label{fig:circle:u_Nh_comp}
  \end{subfigure}
  \caption{(a). The error of pressure on the Eulerian grids. (b). The error of velocity on the Lagrangian grids. (c). The error of velocity on the Eulerian grids. The blue lines with markers `o' are for the `normal' case. The red lines with markers `+' are for the `tangent' case. The solid lines are for the new FSIB method. The dashed lines are for the standard IB method for comparison. The invisible points below the y-axis limit, which are (a) the red solid line, (b) \& (c) the blue solid line, $N<10^2$, are all machine errors, i.e. $1e-16$.}\label{fig:circle:error_Nh}
\end{figure}

First, we need to decide how we should choose the Fourier resolution $L/N$ according to Lagrangian meshwidth $h_b=\mathcal{L}(\Gamma)/N_b$, where $\mathcal{L}(\Gamma)$ is the length of the immersed boundary. As the standard IB method suggests, they should be linearly coupled, so hopefully, we will get the same coupling relation. We fix $N_b=2^6$ and increase $N$ by a power of 2 from $N=2^2$ to $N=2^{10}$ and plot the error of pressure and velocity in \cref{fig:circle:error_Nh}. Generally, there are two types of solutions shown here. One is non-trivial solutions including the pressure for the `normal' case (blue lines with markers `o' in \cref{fig:circle:p_Nh_comp}) and the velocity for the `tangent' case (red lines with markers `+' in \cref{fig:circle:u_Nh_comp_bd,fig:circle:u_Nh_comp}). The others are trivial solutions, i.e. zero solutions, including the velocity for the `normal' case (red lines with markers `+' in \cref{fig:circle:p_Nh_comp}) and the pressure for the `tangent' case (blue lines with markers `o' in \cref{fig:circle:u_Nh_comp_bd,fig:circle:u_Nh_comp}). For the non-trivial solutions, our new FSIB method outperforms the standard IB method by an order of magnitude but still has the same first-order convergence before a transition point. After the transition point, both methods start to accumulate errors. It should be noticed that the transition point is where the Fourier resolution $h=L/N$ and Lagrangian grids $h_b=2\pi r/N_b$ are approximately equal, i.e. $h_b\approx h$. For the trivial solutions, numerical solutions are purely errors and provide us a proxy to check how errors accumulate. We observe that the new FSIB method has machine error before the transition point, which is plausible, and the standard IB method has decreasing errors that are comparable to that of non-trivial solutions. After the transition point, errors of both methods increase to the same level.

Before the transition point, the error decreases because the truncation error in Fourier space decreases as we use more Fourier modes. After the transition point, the boundary effectively develops holes, since the effective width of the regularized delta function becomes less than the distance between the Lagrangian boundary markers. It is interesting and instructive that this occurs for the FSIB method as well as for the standard IB method, since the transition illustrates the somewhat hidden role that the `$\mathtt{sinc}$' kernel is playing in the FSIB methodology.

Thus, in conclusion, we should couple the Fourier resolution and the Lagrangian meshwidth by the condition that $h_b\approx L/N$ which means they should match. In this way, they are linearly coupled just like the standard IB method. Moreover, it is not wise to make the Fourier resolution on the Eulerian grids higher than the Lagrangian grid resolution because it will destroy the machine accuracy of the FSIB method for trivial solutions.

\subsubsection{Spatial convergence of the FSIB method for the Stokes equations.}\label{sec:FSIB_Stokes_conv_2D}

\begin{figure}[!htbp]
  \centering
  \includegraphics[width=.4\textwidth]{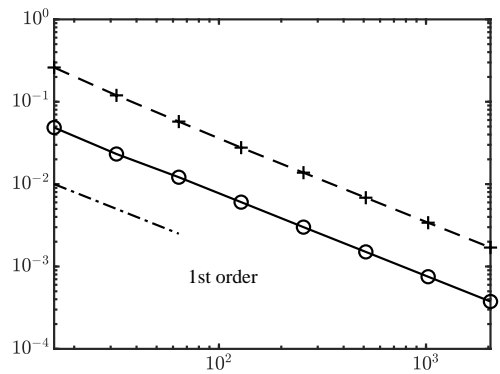}
    \caption{Error of velocity on the Lagrangian grids for the `tangent' case vs $N$ in log-log scale. Lagrangian grid size $N_b$ and the number of Fourier modes $N$ increase simultaneously. The dash-dotted straight line is the first-order reference line. The solid lines with markers `o' are for the FSIB method. The dashed lines with markers `+' are for the standard IB method. }\label{fig:u_Nh_h_comp}
\end{figure}

Given the conclusion above, we couple the Fourier resolution and Lagrangian meshwidth and, in two dimensions, it is equivalent to increasing $N$ and $N_b$ simultaneously to see the real spatial convergence rate of the FSIB method. We start with $N=8$ and $N_b=4$ and double them for each refinement. We plot the successive empirical error of the velocity on the Lagrangian grids for the `tangent' case in \cref{fig:u_Nh_h_comp}, which is the most important non-trivial solution, for 8 successive refinements in total. The standard IB method shows the first-order convergence which is not a surprise and has been addressed for the Stokes equations in \citep{moriConvergenceProofVelocity2008}. The FSIB method shows the same first-order convergence as the standard IB method but outperforms it by an order of magnitude for the Stokes problem in two dimensions.

\subsubsection{Boundary resolution in Stokes flow}\label{sec:bdr_reso_stokes}

\begin{figure}[!htbp]
    \centering
    \includegraphics[width=.7\textwidth]{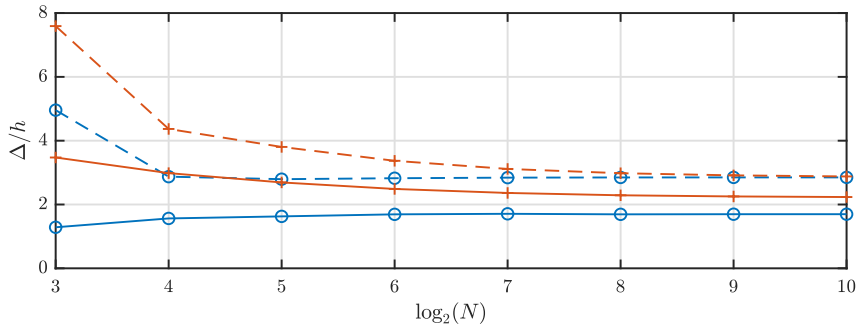}
    \caption{$\Delta/h$ vs $\log_2(N)$, where $\Delta$ is the width of the numerical solutions on the jump discontinuity. Blue lines with markers `o' are for the pressure $p$ on the vertical centerline for the `normal' case, and red lines with markers `+' are for the velocity's derivatives $du_x/dy$ on the vertical centerline for the `tangent' case. Solid lines are for the FSIB method, dashed lines are for the standard IB method.}
    \label{fig:reso_stokes}
\end{figure}

Although we compare the `$\mathtt{sinc}$' kernel with those used in the standard IB method in \cref{fig:kernels_comp} and claim the FSIB method has sharper boundary resolution, we have not quantified the resolution yet. Analytically, the jump continuity has zero width, but numerical solutions always have a finite width, which provides us a way to measure the boundary resolution. On the vertical centerline, the pressure for the `normal' case and the velocity's derivative $du_x/dy$ have jump discontinuity on the boundary. We fit the slope of the jump and divide the height of the jump by the slope to get the width of the jump, denoted as $\Delta$. We increase grid size $N_b$ and $N$ simultaneously as the coupling rule and plot $\Delta/h$ vs $\log_2(N)$ in \cref{fig:reso_stokes}. As the grid refines by two times each time, $\Delta/h$ converges to fixed values for the FSIB method and the standard IB method. That suggests the ratio $\Delta/h$ is an intrinsic property of the FSIB method and the standard IB method no matter how much you refine the grids. For the standard IB method, the boundary resolution converges to $3h$, i.e. three Eulerian meshwidths. For the FSIB method, the boundary resolution converges to about $2h$, which is two thirds of the standard IB method. This shows that the FSIB method has improved boundary resolution for a singular force layer in Stokes flow and we will verify this again in Navier-Stokes flow in \cref{sec:poiseuille}.

\subsection{Navier-Stokes equations}

In this section, we switch from the Stokes equations to the Navier-Stokes equations which involve nonlinearity and time dependence. We start with numerical experiments in two dimensions and then extend them to three-dimensional experiments. 

\subsubsection{Circle with sine velocity in Navier-Stokes flow in two dimensions}\label{sec:circle_2D_NS}

We first consider an initially circular boundary immersed in Navier-Stokes flow, which is a classic problem \citep{linSolvabilityStokesImmersed2019,moriWellPosednessGlobalBehavior2019}. The circle with radius $r=L/4$ is placed at the center $[L/2,L/2]$ of the periodic box $\Omega_L=[0,L]^2$. The immersed boundary is uniformly discretized as $X(\theta_i)=[L/2+(L\cos(\theta_i))/4,L/2+(L\sin(\theta_i))/4]$, where $\theta_i=i\Delta \theta = i\cdot 2\pi/N_b$ and $i=0,...,N_b-1$. The Fourier modes and temporal grids are the same as \cref{sec:discretization}. Note that we should keep $\Delta t \sim h$ for the Runge-Kutta temporal solver to be stable.

The circle is initialized with a perturbation velocity in the y direction as a sine function $\bs u_0(x,y)= [0,\sin(2\pi x/L)]$. And the force density on the immersed boundary is simply given by elasticity

\begin{equation}\label{eq:elstic_force_2d}
\bs F(\theta_i) = K \frac{\bs X(\theta_{i+1})+\bs X(\theta_{i-1})-2\bs X(\theta_i)}{\Delta \theta^2} .
\end{equation}

\begin{figure}[!htbp]
  \centering
  \begin{subfigure}[b]{.95\textwidth}
  \includegraphics[width=\textwidth]{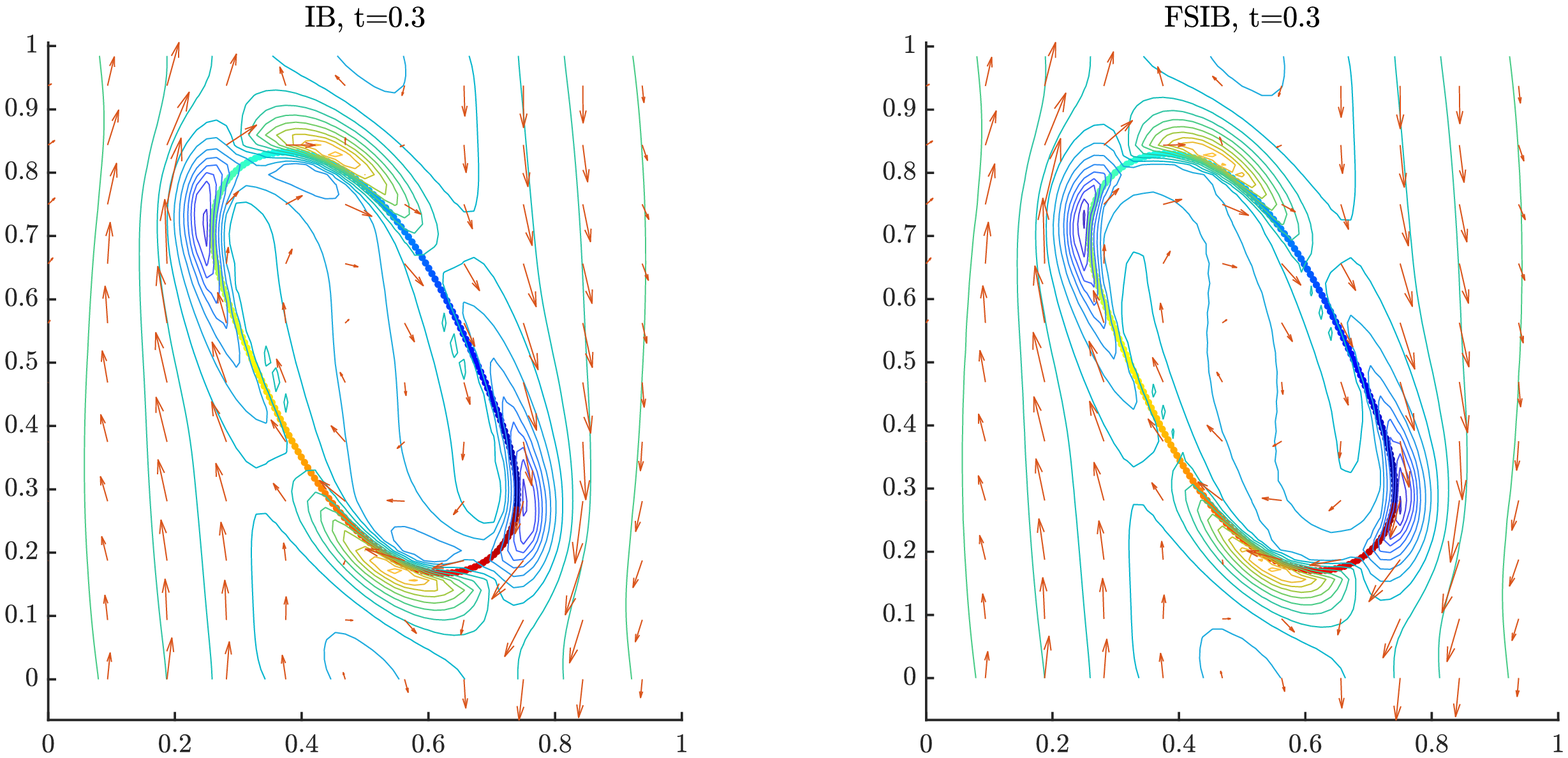}
  \end{subfigure}
  \vfill
  \begin{subfigure}[b]{.95\textwidth}
  \includegraphics[width=\textwidth]{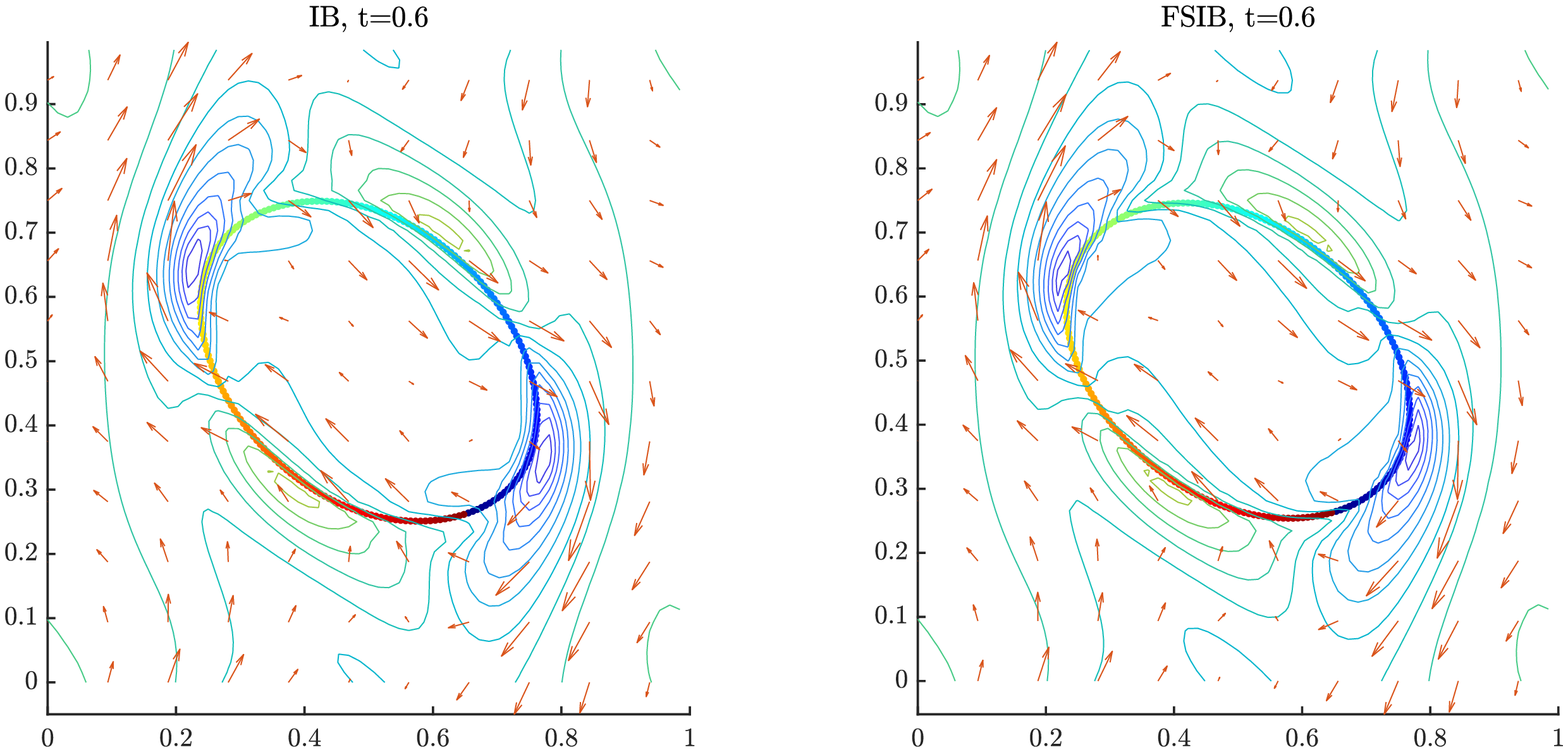}
  \end{subfigure}
  \vfill
  \begin{subfigure}[b]{.95\textwidth}
  \includegraphics[width=\textwidth]{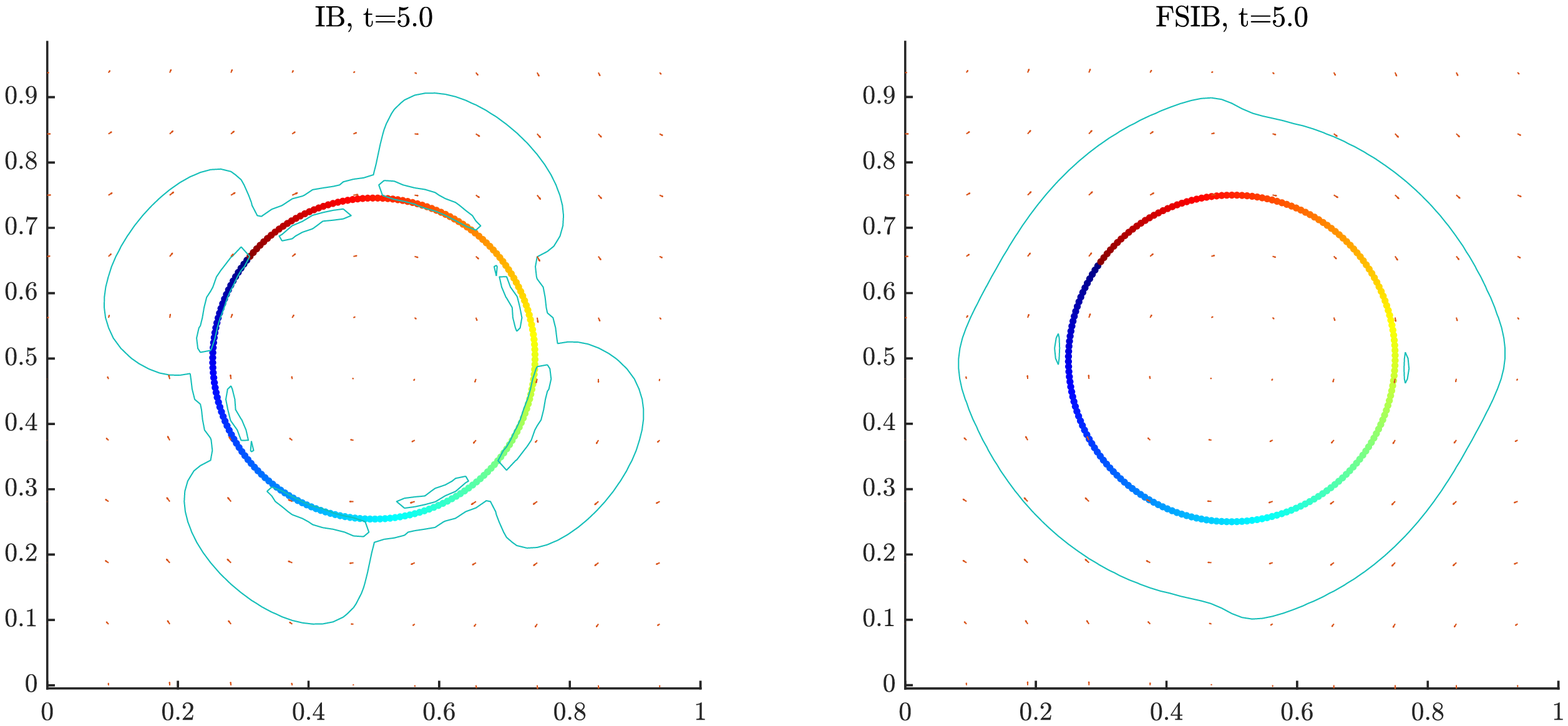}
  \end{subfigure}
  \caption{Snapshots for a circle in Navier-Stokes flow in 2D. The red vectors represent the velocity field with fixed scaling. The contours are for the vorticity with the colormap scaled by a fixed interval. Colors on the immersed boundary encode the Lagrangian coordinate, so a given color stays with a corresponding material segment of the immersed boundary.  Note that the material points of the immersed boundary are rotating clockwise even though there is a wave of deformation at early times that is propagating counterclockwise. The first row is at $t=0.3$, the second row is at $t=0.6$, and the third row is at $t=5.0$ after the immersed boundary returns to a circular shape and is nearly static. The left is for the IB method and the right is for the FSIB method.}
    \label{fig:snapshot_t32}
\end{figure}

\begin{figure}[!htbp]
  \centering
  \begin{subfigure}[b]{0.45\textwidth}
  \includegraphics[width=\textwidth]{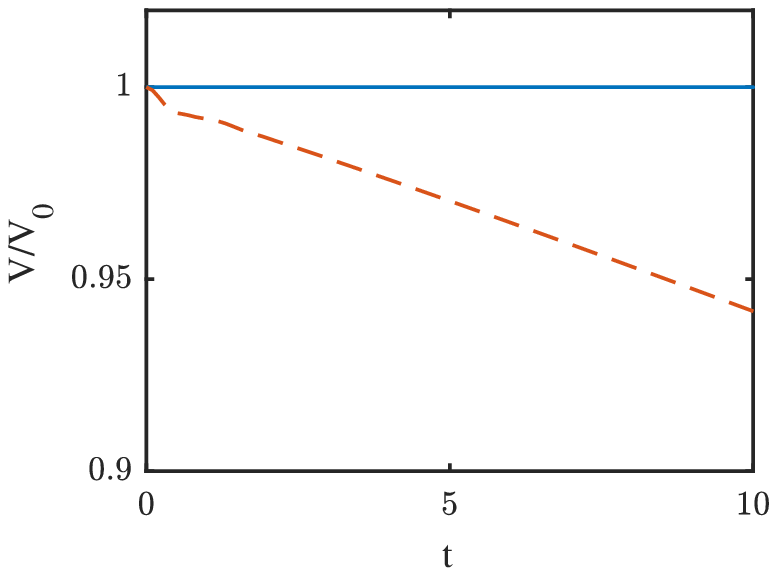}
    \caption{}\label{fig:NS2Dvolume}
  \end{subfigure}
  \hfill
  \begin{subfigure}[b]{0.45\textwidth}
  \includegraphics[width=\textwidth]{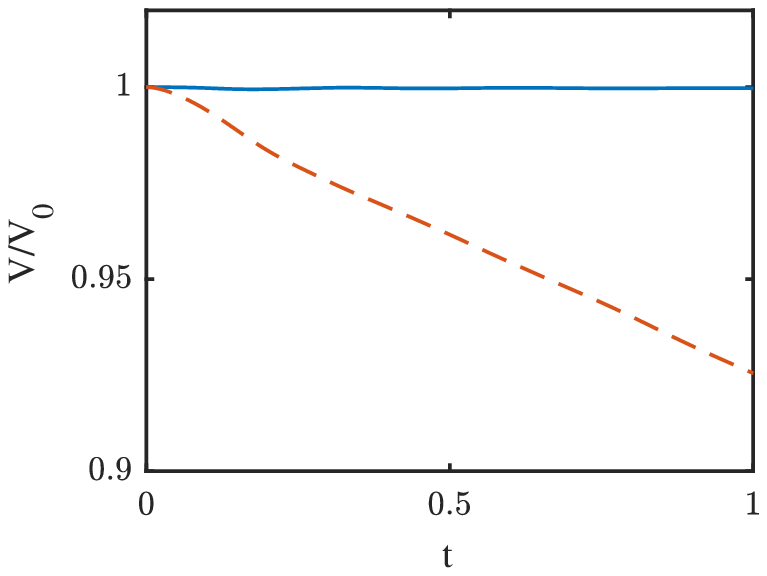}
    \caption{}\label{fig:NS3DElasticvolume}
  \end{subfigure}
  \caption{Normalized volume $V/V_0$ vs t, where $V$ is the volume enclosed by the immersed boundary, $V_0=V(t=0)$. (a). The sine velocity circle problem in Navier-Stokes flow in 2D. (b). The sphere with elastic surface energy in Navier-Stokes flow in 3D. The blue solid lines are for the FSIB method. The red dashed lines are for the standard IB method.}\label{fig:volume_conv}
\end{figure}

First, we run a simple simulation with $L=1$, $N=64$, $N_b=202$ such that $h_b\approx L/N$, $\Delta t=1e-3$, $K=1$, $\rho=1$, $\mu=0.01$. The total simulation time $T=10$ is long enough that we observe that the immersed boundary returns to its initial circular geometry, and we check how the volume enclosed by the immersed boundary evolves. We compared the results of our new FSIB method with the standard IB method in a movie \footnote{\url{https://www.math.nyu.edu/~zc1291/public/IBM/NUFFT/comp_N_64_spec_ns_T_32_colored.mp4}} and also plot representative snapshots with fluid velocity fields and vorticity contours in \cref{fig:snapshot_t32}. In the beginning, the sine perturbation flow drives the circle to deform and the two methods show consistency. Then, the oscillation decays, the flow starts to settle down, and the immersed boundary changes back to a circle. The discrepancy between the solutions of the IB method and the FSIB method grows with time. The IB method is more dissipative and develops defects of the vorticity contours near the boundary after the boundary returns to a circular configuration. Moreover, the immersed boundary force law, \cref{eq:elstic_force_2d}, that we are using would make the immersed boundary shrink to a point if this were not prevented by the incompressibility of the fluid. As the movie evolves, the volume of the circle in the IB method shrinks gradually and this effect is large enough to be visible in the movie. The volume leakage of the IB method can also be noticed by the snapshots in \cref{fig:snapshot_t32} and the circle in the FSIB method remains the same volume from the beginning to the end. We can quantify the change of volume by calculating the area enclosed by the immersed boundary numerically, denoted by $V(t)$, at each snapshot for both methods and plot $V(t)/V_0$, where $V_0$ is the initial volume, as a function of $t$ in \cref{fig:NS2Dvolume}. It shows that the new FSIB method has volume conservation with no leakage at all within numerical accuracy, while the standard IB method leaks volume linearly in time.

\begin{figure}[!htbp]
  \centering
  \begin{subfigure}[b]{0.45\textwidth}
  \includegraphics[width=\textwidth]{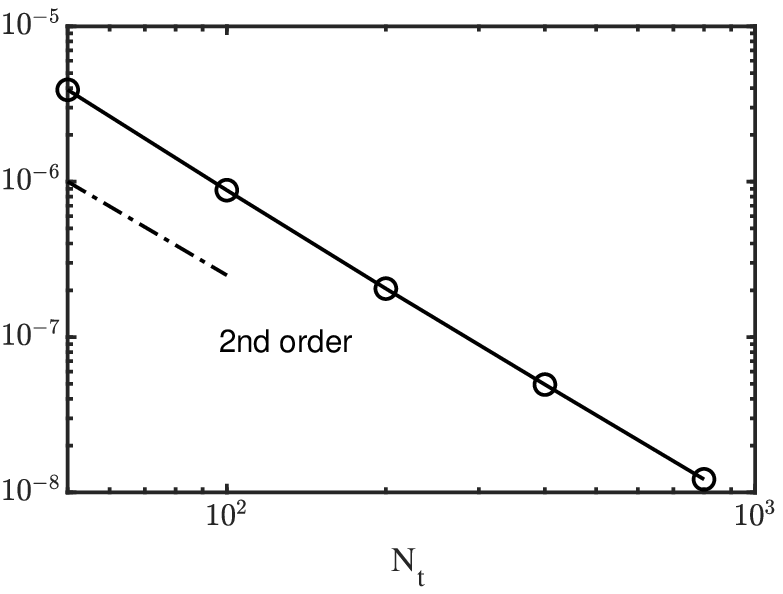}
    \caption{Empirical relative norm-2 error of velocity at the final time $t=0.1$ for different refinements of temporal grids for a circle in Navier-Stokes flow in 2D. The solid line with markers `o' is for the FSIB method and the dash-dotted line is the reference for the second-order convergence.}\label{fig:NS2DerrNt}
  \end{subfigure}
  \hfill
  \begin{subfigure}[b]{0.45\textwidth}
  \includegraphics[width=\textwidth]{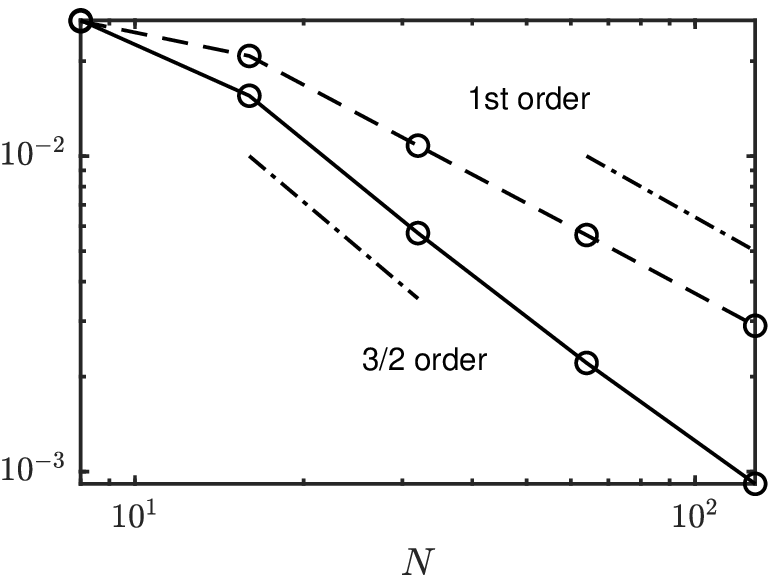}
    \caption{Empirical relative norm-2 error of velocity at the final time $t=0.1$ for refinements of both spatial and temporal grids for a circle in Navier-Stokes flow in 2D. The solid line is for the FSIB method and the dashed line is for the IB method. The dash-dotted lines are for reference (3/2 order convergence for the FSIB and 1st order for the IB).}\label{fig:spatial_conv}
    \end{subfigure}
\end{figure}

Also, we verify the temporal second-order convergence of the FSIB method since it is a Runge-Kutta-2 integrator. We start with the largest stable time step $\Delta t=0.002$ and refine the temporal grid by a factor of two every time. Then the relative error of the solution at the elapsed time $t=0.1$ of different time resolutions is plotted in \cref{fig:NS2DerrNt}. It shows a perfect second-order convergence that is consistent with the theory. Note, however, that only the time step was refined in this study. The number of Fourier modes and the number of Lagrangian markers on the immersed boundary were kept constant.

We also couple the spatial grids and the temporal grids to check the overall convergence rate. We start with $N=8$, $\Delta t=0.001$, $N_b=26$ and, for each refinement, we halve $\Delta t$ and double $N$ and $N_b$. The relative error of the velocity at $t=T=0.1$ is plotted in \cref{fig:spatial_conv}. The overall convergence rate is first-order for the IB method, but the new FSIB method outperforms by half an order in this two-dimensional Navier-Stokes problem.

\subsubsection{Poiseuille flow governed by Navier-Stokes equations in a two-dimensional channel}\label{sec:poiseuille}

Consider the classic problem of the Navier-Stokes flow passing through a Poiseuille channel in two dimensions. The channel walls are placed at $y=0$ and $y=D$. We have the no-slip boundary condition on the channel walls so $\bs u(x,y=0,t)=\bs u(x,y=D,t)=0$. The flow is initialized at rest, i.e. $\bs u(x,y,t=0)=0$. A uniform force field $f_0$ in the x direction is applied as $\bs f(x,y,t)=(f_0,0)$. One of the reasons that we choose this problem is that, for comparison with our numerical results, we can derive the analytical solutions as follows. In the range of laminar flow, the velocity field is simply in the x direction as a function of $y$, so $\bs u=(u(y,t),0)$, which allows us to simplify the Navier-Stokes \cref{eq:navierstokes} as an equation for $u(y,t)$

\begin{equation}\label{eq:poiseuille}
u_t=\frac{\mu}{\rho} u_{yy}+\frac{f_0}{\rho}
\end{equation}
with the no-slip boundary condition $u(0,t)=u(D,t)=0$. So we can write the solution as a linear combination of eigenfunctions 

\begin{equation}
u(y,t)=\sum_{k=1}^{\infty}u_k(t)\sin(\lambda_k y),
\end{equation}
where eigenvalues $\lambda_k=\frac{k\pi}{D}$. We solve \cref{eq:poiseuille} as 

\begin{equation}
u_k=\frac{f_k}{\mu \lambda_k^2}(1-\exp(-\frac{\mu}{\rho}\lambda_k^2t)).
\end{equation}
We also write the force field as $f_0=\sum_{k=1}^{\infty}f_k\sin(\lambda_k y)$, where $f_k=\frac{2}{D}\int_0^D f_0\sin(\lambda_k y) dy$. This gives us

\begin{equation}
\left\{
\begin{array}{ll}
     f_{2k+1}&=\frac{4f_0}{(2k+1)\pi},  \\
     f_{2k}&=0 ,
\end{array}
\right. \quad k\ge 0.
\end{equation}
Therefore, we have the analytical solution to the Poiseuille flow as 

\begin{equation}\label{eq:poiseuille_velo}
u(y,t)=\sum_{k=0}^{\infty} \frac{4f_0}{\mu \lambda_{2k+1}^2(2k+1)\pi}\left(1-\exp(-\frac{\mu}{\rho}\lambda_{2k+1}^2 t)\right)\sin(\lambda_{2k+1}y).
\end{equation}
The flux through the Poiseuille channel $\Phi(t)$ is then given by 

\begin{equation}\label{eq:poiseuille_flux}
\Phi(t)=\int_0^D u(y,t)dy=\sum_{k=0}^{\infty} \frac{8f_0}{\mu \lambda_{2k+1}^3(2k+1)\pi}\left(1-\exp(-\frac{\mu}{\rho}\lambda_{2k+1}^2 t)\right).
\end{equation}

It is worth noticing that as $t\xrightarrow[]{}\infty$, the fluid goes to equilibrium and the sine series in \cref{eq:poiseuille_velo} converges to the famous steady state solution of the Poiseuille flow as 

\begin{equation}\label{eq:steady_poiseuille_velo}
u_{eq}(y)=\frac{f_0}{2\mu}y(D-y)
\end{equation}
with steady flux

\begin{equation}\label{eq:steady_poiseuille_flux}
\Phi_{eq}=\frac{f_0D^3}{12\mu}.
\end{equation}

We implement the FSIB method as follows. The Poiseuille channel is placed at the center of a periodic box $\Omega_L=[0,L]^2$ with the boundary at $y=L/4$ and $y=3L/4$, so the width of the channel is $D=L/2$. The Lagrangian meshwidth and the Fourier resolution are coupled as $L/N=h_b$. The no-slip boundary condition is implemented by the target point method as two static walls. The target point method introduces a fixed point that is called the target point for each Lagrangian marker on the wall and uses a spring force to keep each Lagrangian marker close to its target position. The idea of this method dates back to the original paper of the IB method to simulate a heart valve \citep{peskinNumericalAnalysisBlood1977} where static structures to which moving valve leaflets are attached are implemented by the target point method. The target point method was later developed into a complete theory to simulate moving or static immersed boundary with or without the additional mass \citep{kimPenaltyImmersedBoundary2016}. Most recently, the numerical stability of the target point method has been analyzed in \citep{huaAnalysisNumericalStability2022}. Here we put fixed target points on the boundary at $\tilde{\bs X}_i=(ih,L/4),  \tilde{\bs X}_{i+N}=(ih,3L/4),\  i=0,...,N-1$, so that $N_b=2 N$. The Lagrangian points are initialized on the fixed target points as $\bs X_i(t=0)=\tilde{\bs X}_i$. The Lagrangian points are attached to the target points on the boundary by the spring force

\begin{equation}
\bs F_i(t) = K (\tilde{\bs X_i}-\bs X_i(t)),\ i=0,...,N_b-1.
\end{equation}
This model has fixed boundary walls unlike the freely moving boundary in \cref{sec:circle_2D_NS}. Moreover, it will provide us a way to quantitatively measure the effective width of the boundary.

\begin{figure}[!htbp]
  \centering
  \begin{minipage}[b]{0.45\textwidth}
  \includegraphics[width=\textwidth]{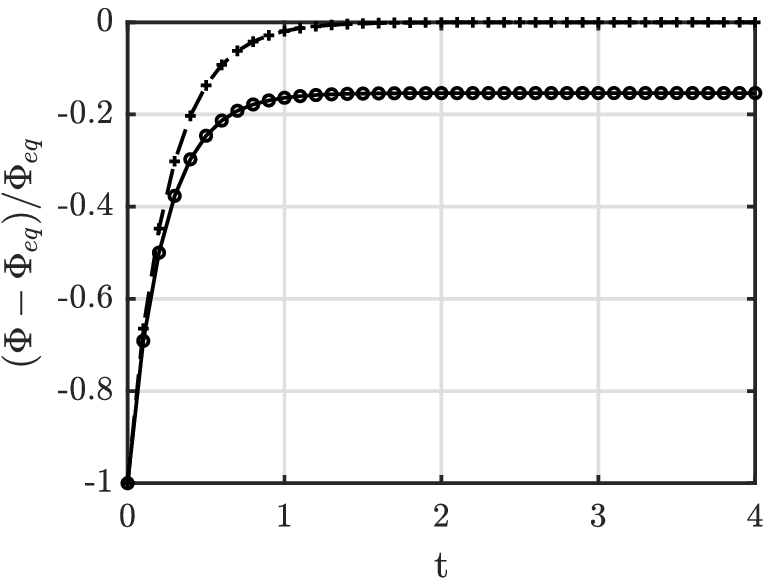}
    \caption{Numerical flux (solid line with marker `o') on the coarse grid $N=8$ compared with analytical flux (dashed line with marker `+') as functions of time}\label{fig:flux_comp}
  \end{minipage}
  \hfill
  \begin{minipage}[b]{0.45\textwidth}
  \includegraphics[width=\textwidth]{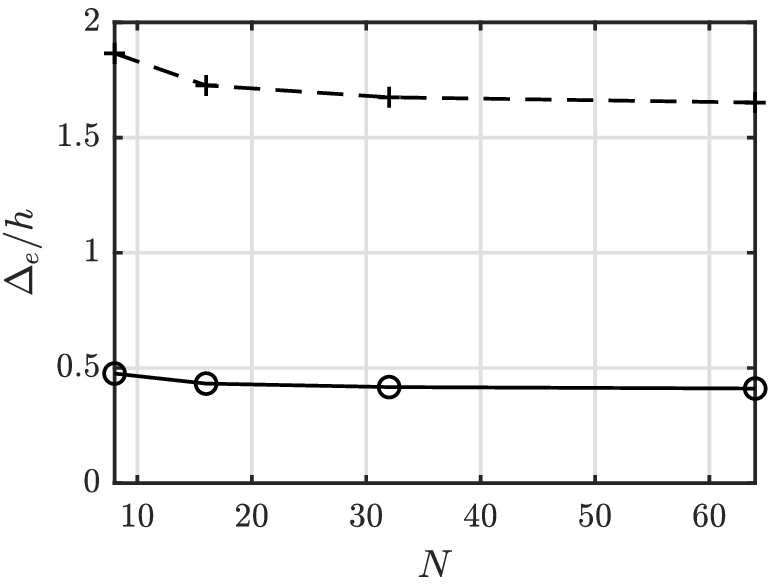}
    \caption{Effective width of the boundary with normalization $\Delta_e/h$ vs $N$. The solid line marked by `o' is the FSIB method, and the dashed line marked by `+' is the standard IB method}\label{fig:effective_width}
  \end{minipage}
\end{figure}

We set $L=1,\ \mu=0.1,\ \rho=1,\ f_0=100$ and start with a coarse grid where $N=8$. First, we compute the flux of the numerical solution of the FSIB method as a function of time until equilibrium and compare it with the analytical flux in \cref{eq:poiseuille_flux}. We plot $(\Phi(t)-\Phi_{eq})/\Phi_{eq}$ vs $t$ in \cref{fig:flux_comp}. The analytical result starts from $-1$, i.e. zero flux, and converges to $0$ as $t\rightarrow \infty$. The numerical result has a similar evolution but converges to a smaller value than the analytical solution. This means the implementation of the boundary has the effect to slow the flow as a result of the finite resolution. Thus, it provides us a way to quantify the boundary resolution. From the steady-state flux formula in \cref{eq:steady_poiseuille_flux}, we know the effective width of the channel for a given equilibrium flux $\Phi_{eq}$ is

\begin{equation}
D_e = \left(\frac{12\mu\Phi_{eq}}{f_0}\right)^{1/3} .
\end{equation}
Therefore, $\Delta_e=D-D_e$ is a good measurement of the resolution of the boundary, which we call the effective boundary width. We compare the effective width of the boundary of the FSIB method and the standard IB method. We plot $\Delta_e/h$ vs $N$ in \cref{fig:effective_width} and observe that $\Delta_e/h$ converges to fixed values for both methods, which is similar to the Stokes flow case in \cref{sec:bdr_reso_stokes}. For the standard IB method, the boundary width converges approximately to $1.7h$. For the FSIB method, the boundary width converges to approximately $0.4h$, which is about $4$ times smaller than the standard IB method. This result quantifies the improvement in boundary resolution achieved by the FSIB method.

\begin{figure}[!htbp]
  \centering
  \begin{subfigure}[b]{0.45\textwidth}
  \includegraphics[width=\textwidth]{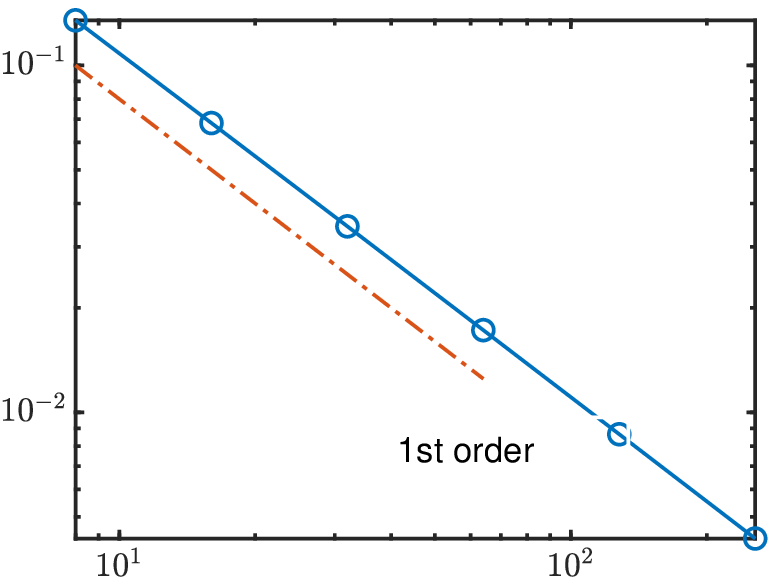}
    \caption{}\label{fig:err_2_steady_poiseuiile}
  \end{subfigure}
  \hfill
  \begin{subfigure}[b]{0.45\textwidth}
  \includegraphics[width=\textwidth]{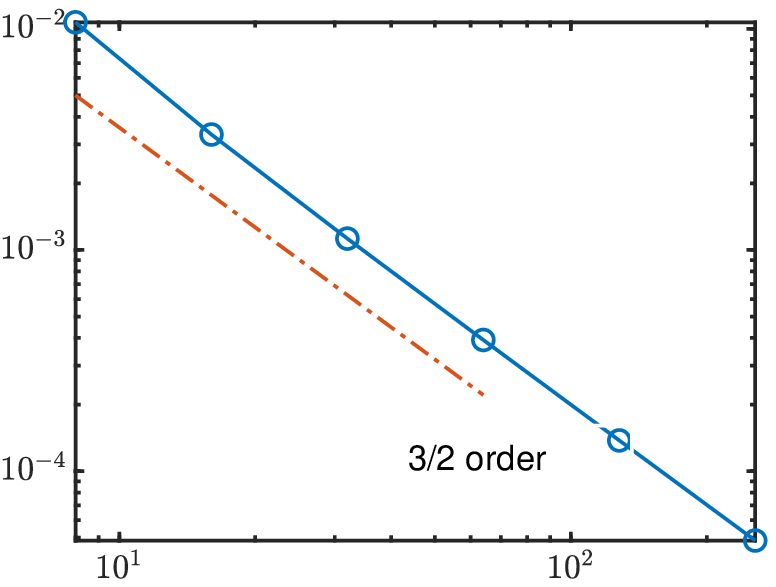}
    \caption{ }\label{fig:err_2_steady_poiseuiile_3_halves_effective_width}
    \end{subfigure}
    \caption{Relative error in norm 2 of the velocity field in the channel at equilibrium vs $N$ in log-log scale. Compare the solution of the FSIB method with the analytical solution given by channel width (a) $D$  (b) $D_e$. The blue solid lines with markers `o' are for the FSIB method and the red dash-dotted lines are for reference (1st order for the left and 3/2 order for the right).}
\end{figure}

The fixed boundary affects the convergence rate of the FSIB method. For simplicity, we set the channel width $D=L$ so only one boundary wall at $y=0$ is needed due to periodicity, and we only solve the equilibrium problem, which is governed by the Stokes equations. We start with $N=N_b=8$ and refine the grids 6 times and plot the relative errors of the velocity field compared with analytical solutions in \cref{fig:err_2_steady_poiseuiile}. The FSIB method shows first-order convergence, which is no surprise because it is the same as the convergence rate of the previous Stokes problem in \cref{sec:FSIB_Stokes_conv_2D}. If we compare the numerical solution with the analytical solution with the effective channel width, i.e. replace $D$ by $D_e$ in \cref{eq:steady_poiseuille_velo}, however, we see $3/2$ order convergence in \cref{fig:err_2_steady_poiseuiile_3_halves_effective_width}. This verifies that the static boundary is a first-order effect that compromises the convergence rate of the FSIB method and the effective width of the channel represents the boundary effect.

\subsubsection{Elastic Sphere in Navier-Stokes flow in three dimensions}\label{sec:sphere_energ}

\begin{figure}[!htbp]
  \centering
  \begin{subfigure}[b]{\textwidth}
  \includegraphics[width=\textwidth]{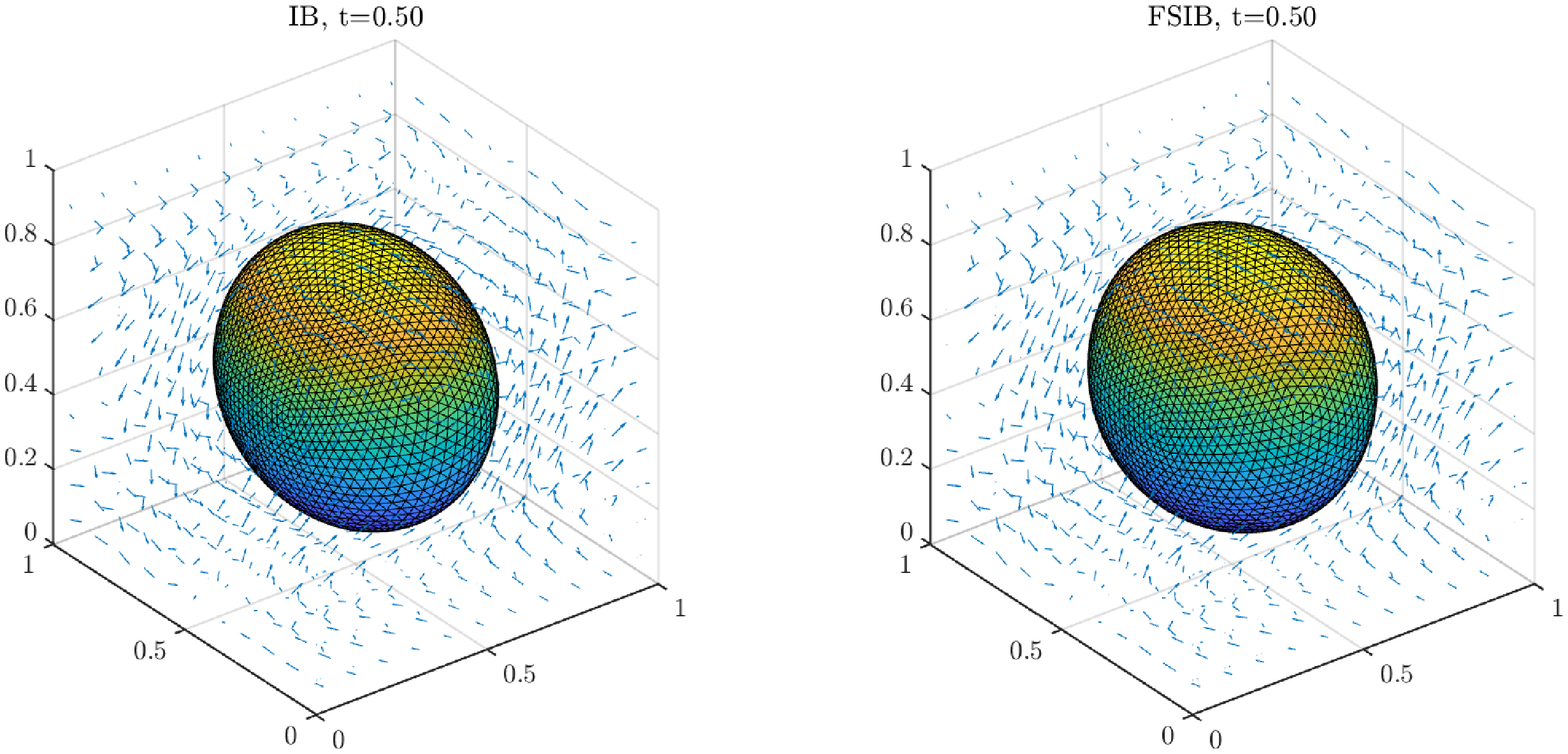}
  \end{subfigure}
  \vfill
  \begin{subfigure}[b]{\textwidth}
  \includegraphics[width=\textwidth]{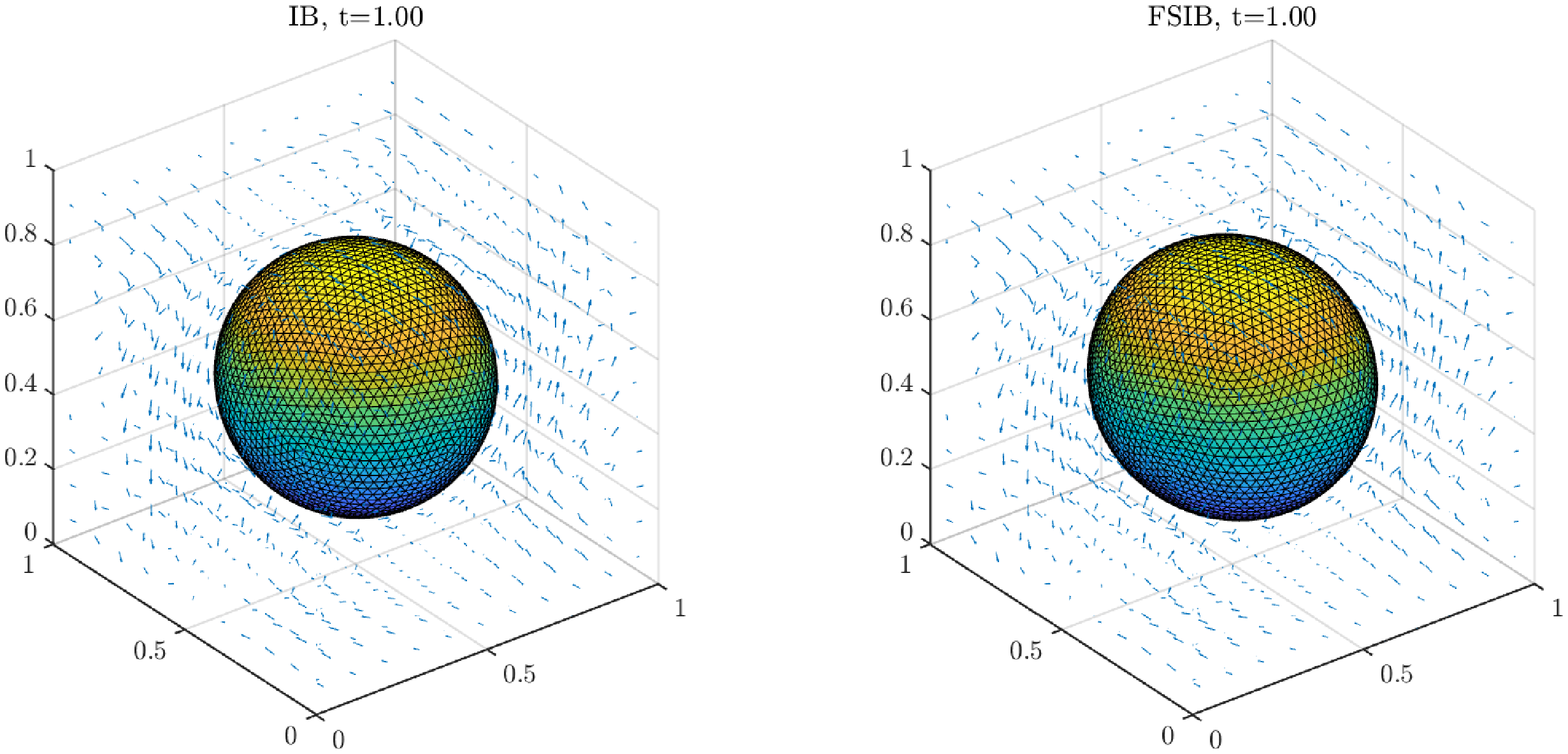}
  \end{subfigure}
    \caption{Snapshots for the elastic sphere in Navier-Stokes flow in 3D. The blue vectors represent fluid velocity fields. The top is for the intermediate state when $t=0.5$ and the bottom is for the final state when $t=1.0$. The left is for the IB method and the right is for the FSIB method.}
    \label{fig:snapshot_t1}
\end{figure}

Now we extend numerical experiments to three dimensions. The analog of a circle in three dimensions is a sphere. So, we start with the most simple case as a sphere with radius $r=L/3$ that is placed at the center of a periodic box $\Omega_L = [0,L]^3$. But a uniform discretization for a sphere is not possible while it is easy to obtain for a circle in two dimensions. Thus, we turn to triangulation on the sphere and use the nodes of triangles as the discrete point $\bs X_i$, $i=0,...,N_b-1$. The edge length of triangles should be as uniform as possible. The regular icosahedron with 12 vertices provides a beautiful starting point for the triangulation. Successive refinement will provide any Lagrangian resolution that is required. For each refinement, we split each triangle into four smaller triangles by adding new points on the midpoints of the three edges and projecting the three midpoints to the spherical surface to get the new vertices and triangles. The edge length is approximately halved and the number of triangles is four times greater. Here we visualize the triangulation with the refinement of level four in \cref{fig:snapshot_t1} and one can refer to fig. 11 in \citep{baoImmersedBoundaryMethod2017} for detailed visualization of such a triangulation. Although the triangulation looks regular, it is important to note that this is not quite the case.  In particular, the refinement process produces triangles with a variety of shapes, although all of them are close to equilateral.

The continuous elastic force field on the immersed boundary is modeled by springs on the edges of the triangles. So, the force at any node $X_i$ is given by all the edges attached to this node as

$$\bs F_i = \sum_{j \text{ linked with }i} K_{i,j}(\bs X_j- \bs X_i),$$
where $K_{i,j}$ is the spring constant of the edge that links node $\bs X_i$ and $\bs X_j$. The choice of $K_{i,j}$ that immediately comes to one's mind might be a constant $K_{i,j}$, but the fact is that this is not a physical model, i.e., it has no continuum limit, even if the constant K is made a function of the refinement level in an attempt to produce such a continuum limit. Therefore, the choice of constant $K_{i,j}$ should be avoided. To derive a physically reasonable model we consider the elastic energy density

$$\epsilon_0(t)=\frac{\sigma}{2}(\lambda_1^2(t)+\lambda_2^2(t)),$$
where the subscript $0$ means this energy density should be integrated over the spherical reference configuration. The constant $\sigma$ represents the strength of elasticity. The variables $\lambda_1$ and $\lambda_2$ are the principal stretch ratios at time $t$. The elastic energy minimizes as the immersed boundary shrinks to a point but the incompressible fluid prevents this from happening. So this is an ideal experiment to verify the volume conservation of the FSIB method. For any triangle $\Delta ABC$ of the triangulation, we denote the edge lengths as $a=|\bs{BC}|$, $b=|\bs{AC}|$, $c=|\bs{AB}|$. The initial reference triangle edges at $t=0$ are thus denoted as $a_0,b_0,c_0$. So, at any time $t$, the stretch on the triangle $\Delta ABC$ is calculated as 

$$\lambda_1^2+\lambda_2^2=\frac{\sum_{a,b,c}a^2(b_0^2+c_0^2-a_0^2)}{8 S_0^2(\Delta ABC)},$$
where the subscript $a,b,c$ represents the circular summation of any permutation of $a,b,c$ and the $S_0^2(\Delta ABC)$ is the area of the triangle at $t=0$. Now multiply the force density by the reference triangle area $S_0$ and sum over all of the triangles. This gives the total energy on the sphere as

\begin{equation}\label{eq:total_elastic_energy}
E=\frac{\sigma}{2}\sum_{\Delta ABC}\frac{\sum_{a,b,c}a^2(b_0^2+c_0^2-a_0^2)}{8 S_0(\Delta ABC)}.
\end{equation}
Thus, the tension in the edge $|\bs{AB}|=c$ should be the negative derivative of the energy as
$$T_{AB}=-\frac{\partial E}{\partial c}=-\sigma\frac{\sum_{\Delta_c} c(a_0^2+b_0^2-c_0^2)}{8 S_0(\Delta_c)},$$
where the notation $\Delta_c$ represents the two triangles that edge $c$ attaches to. Note that the tension is given by a linear function of the edge length $c$ and thus satisfies the spring model. Therefore, the spring constant on any edge $\bs{AB}$ is

\begin{equation*}
 K_{\bs{AB}} = \sum_{\Delta_c}\frac{\sigma}{8}\frac{a_0^2+b_0^2-c_0^2}{S_0(\Delta_c)} = \sum_{\Delta_c} \frac{\sigma}{8}\frac{2ab\cos(\angle C)}{\frac{1}{2}ab\sin(\angle C)}=\frac{\sigma}{2}(\cot(\angle C_1)+\cot(\angle C_2)), 
\end{equation*}
where $\angle C$ denotes the angle of node $C$ in the triangle $ABC$ at time $t=0$. $C_1$ and $C_2$ are the nodes opposite edge $\bs{AB}$ in the two triangles that have the edge $\bs{AB}$ in common. Thus, when we initialize the simulation, we can compute the stiffness, i.e. spring constant, of each edge $\bs{AB}$. Then we store these spring coefficients of each edge and use them to compute force at each time step. This model of continuous elastic energy simplifies to springs that connect edges of triangles with various stiffness $K$. 

Another way to compute the discrete force of the elasticity model is to rewrite the total elastic energy in \cref{eq:total_elastic_energy} as a function of the coordinates of all the nodes, i.e. substitute $a=|\bs B-\bs C|$, $b=|\bs A-\bs C|$, $c=|\bs A-\bs B|$. Then take the negative derivative of the total energy with respect to the node coordinate, e.g. $\bs F_{\bs A}=-\partial E/ \partial \bs A$, to obtain the elastic force on the node. This way avoids the introduction of the spring model but is less intuitive. 

At time $t=0$, the flow is initialized as a sine wave $\bs u(x,y,z)=[\sin(2\pi z/L),0,\sin(2\pi y/L)]$, which is similar to what was done in \cref{sec:circle_2D_NS}. Here we first present a numerical experiment on this problem. We set parameters $L=1$, $N=32$, $\rho=1$, $\mu=0.01$, $T=1$, $\Delta t=0.001$ and $\sigma=1$. For the discretization on the sphere, we do four steps of refinements of the regular icosahedron as described above such that the Lagrangian meshwidth matches with the Fourier resolution. We visualize the results of the standard IB method in comparison with our new FSIB method as a movie \footnote{\url{https://www.math.nyu.edu/~zc1291/public/IBM/NUFFT/comp_N_32_spec_ns_Elastic.mp4}} and also the snapshots in \cref{fig:snapshot_t1}. The FSIB method shows a consistent solution with the IB method. The simulation ends at $t=1$, which is enough time for the elastic immersed boundary to return approximately to its initial spherical configuration.

A physical quantity that is worth observing is the volume enclosed by the immersed boundary. To quantify the error in volume conservation, we evaluate the volume enclosed by the triangulated surface, which is given by

\begin{equation*}
    V=\frac{1}{6}\sum_{(i,j,k) \in \texttt{Tri} } (\bs X_j-\bs X_k, \bs X_k-\bs X_i, \bs X_i-\bs X_j) ,
\end{equation*}
where $\texttt{Tri}$ denotes the circular triads of all triangles in the form of three circular numbers $(i,j,k)$ and the circular triple product of three vectors $(\bs a, \bs b,\bs c)$ is defined as $(\bs a,\bs b,\bs c):=\bs a \cdot (\bs b \times \bs c)$.The order of these triads is taken to be counterclockwise when viewed from outside of the surface. It follows that every edge is traversed in opposite directions in the two triangles to which that edge belongs. Note that this formula for the volume enclosed by the triangulated surface gives a result that is independent of the choice of origin, which can even be outside of the surface, and the formula is still correct even if the surface has a complicated shape in which a ray emanating from the origin intersects the surface more than once. These features are consequences of using the \textit{signed} volumes of the individual tetrahedra instead of summing the absolute values of those volumes. So, we can plot the volume that is normalized by the initial volume $V(t)/V_0$ as a function of time in \cref{fig:NS3DElasticvolume}. We have here the same conclusion for the volume conservation in three dimensions as in two dimensions that the new FSIB method has the property of volume conservation while the standard IB method has constant-rate volume leakage as time evolves.

\begin{figure}[!htbp]
  \centering
  \begin{subfigure}[b]{0.45\textwidth}
  \includegraphics[width=\textwidth]{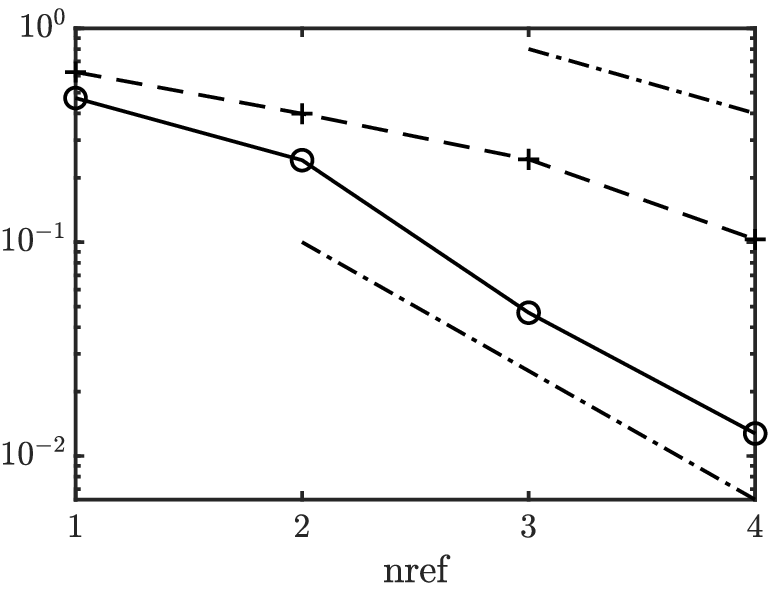}
    \caption{}\label{fig:err_elastic_sphere}
  \end{subfigure}
  \hfill
  \begin{subfigure}[b]{0.45\textwidth}
  \includegraphics[width=\textwidth]{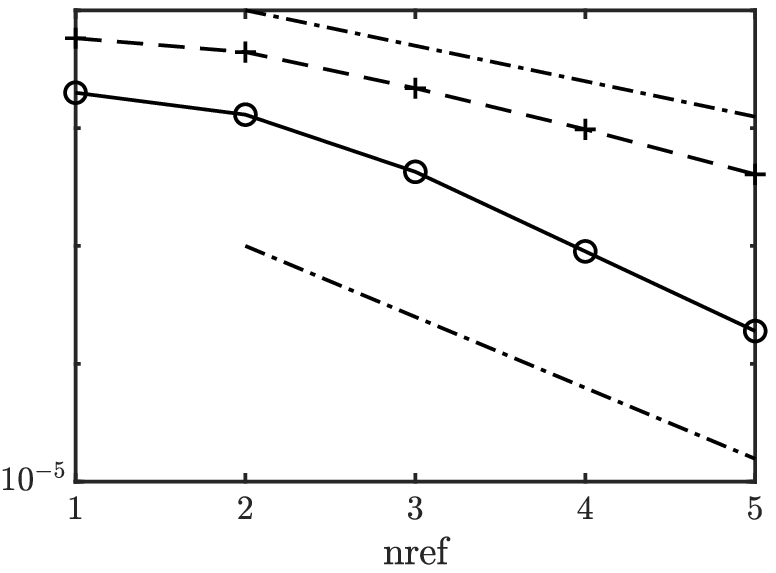}
    \caption{ }\label{fig:err_euler_1st_2nd}
    \end{subfigure}
    \caption{Relative error in norm 2 of the interpolated velocity field at the final snapshot $T=0.2$ in log scale vs the number of refinements. (a) An elastic sphere in Navier-Stokes flow in 3D. (b) A sphere with surface tension in Navier-Stokes flow in 3D. Solid lines with markers `o' are for the FSIB method. Dashed lines with markers `+' are for the IB method. Dash-dotted lines are for reference (second-order convergence for the FSIB method and first-order convergence for the IB method).}
\end{figure}

The main purpose of this experiment on elastic surface energy is to check the convergence of the FSIB method in three dimensions. We start with the coarsest grid where $N=4$, $\Delta t=0.004$, and the number of refinements of the regular icosahedron is $1$. Then, we refine the Lagrangian grids, i.e. the triangulation, on the immersed boundary, halve the time step and double the number of Fourier modes in each direction at the same time for each successive refinement. After that, we compute the empirical relative error in norm 2 of the interpolated velocity field on Eulerian grids at the final time at $t=T$. Due to the scaling of the grids in three dimensions being massive, we managed to simulate only until $T=0.2$ and do the refinement up to $5$ times. We plot the relative error in log scale vs the number of refinement steps in \cref{fig:err_elastic_sphere}. The result is clear that the FSIB method shows second-order convergence while the IB method has only first-order convergence. The advantage of the FSIB method over the IB method in three dimensions regarding the convergence rate is even greater than in the two-dimensional case, which is surprising. Although the standard IB method is generally first-order, there is an IB-like method called the immersed interface method \citep{levequeImmersedInterfaceMethod1994} that has second-order convergence by a modification of the finite difference operator near the immersed boundary. No such modification is needed here.

It is also of interest to consider the case of surface tension. The surface tension of a sphere is modeled by a total energy $E$ that is proportional to the surface area $S$ of the geometry as

$$E=\gamma S,$$
where the constant $\gamma$ is the surface tension and one may easily verify that the total surface energy will equal to the elastic energy in the spherical reference at the initial state configuration in \cref{eq:total_elastic_energy} if we set $\gamma=\sigma$. We make this choice $\gamma=\sigma=1$ so that the stiffness of the two experiments will be similar. Unlike the elastic energy used previously, the surface tension will not prevent the geometry from deformation since the energy only depends on the surface area rather than the geometry. It follows that the surface tension does not oppose changes in the shape of the individual triangles. Therefore, it provides different dynamics. Moreover, the surface tension model produces force in the normal direction to the immersed boundary. This means there is no tangential jump in stress on the immersed boundary. It is also true in this case that the normal derivative of the tangential velocity has no jump across the immersed boundary \citep{laiRemarkJumpConditions2001}.

We use a triangulated surface as before and approximate the surface area as the sum of the areas of the triangles. Following a similar process here as in the case of the elastic model, we take the negative derivative of the energy with respect to the node coordinates and obtain the force on the node $\bs X_i$ given by

$$\bs F_i=-\frac{\gamma}{2}\sum_{(i,j,k)\in \Delta_{i}} (\bs X_j-\bs X_k)\times \bs n_{i,j,k},$$
where $\Delta_i$ denotes all the triangles in the form of the three indexes $(i,j,k)$ of the three vertices to which node $\bs X_i$ is attached, and $\bs n_{i,j,k}$ is the unit normal vector of the triangle $(i,j,k)$ with the positive direction given by the right-hand rule of the ordering $\bs X_i$, $\bs X_j$ $\bs X_k$.

First, we run the essentially same simulation as the elastic model and get similar results for volume conservation. For the convergence study, we start with $N=4$ and one refinement of the regular icosahedron. Time step $\Delta t=0.004$ and we run until $T=0.2$ to compute the interpolated velocity on Eulerian grids. We manage to run 6 times of refinements of the grids and plot the successive relative error in norm 2 of the velocity field in \cref{fig:err_euler_1st_2nd}. The FSIB method shows second-order convergence while the IB method has only first-order convergence, which is the same result as the elastic model. 

\section{Conclusions}

In this paper, we present a new Fourier Spectral Immersed Boundary method. The new method is fully independent of any Eulerian grid and solves the fluid equations in Fourier space. The force spreading and the velocity interpolation of the standard IB method become an integral over the Lagrangian variables and a Fourier series evaluation, respectively, both of which are implemented by the NUFFT. We demonstrate that the FSIB method has an analog in the framework of the IB method by using a new `$\mathtt{sinc}$' kernel. The `$\mathtt{sinc}$' kernel is not finitely supported, as are the standard IB kernels, but it shares many similarities with the standard kernels. It satisfies all the conditions of the standard kernels discretely and continuously and may be viewed as the limit of the standard kernels as the width of their support goes to infinity. We also show that the FSIB method preserves the duality of the force spreading and the velocity interpolation. The conservation of momentum and the conservation of energy are proved. We implement the FSIB method efficiently with the help of the NUFFT with a complexity of $O(N^3\log(N))$ per time step where $N$ is the number of Fourier modes in each dimension. Besides these properties that are shared with the IB method, the FSIB method substantially outperforms the IB method in the following ways. The interpolated velocity field of the FSIB method is analytically divergence-free and thus conserves volume with no leakage. Moreover, the FSIB method has the property of exact translation invariance, and this is impossible to achieve with the finitely supported kernels of the standard IB method. 

We verify these properties of the FSIB method in comparison with the IB method by a series of numerical experiments for the Stokes equations and the Navier-Stokes equations in two space dimensions and also in three space dimensions. The convergence rate of the standard IB method is first-order in all experiments. The FSIB method has the same first-order convergence for Stokes flow but with an empirical error that is 10 times smaller than in the case of the standard IB method. The FSIB method shows $3/2$ order accuracy for the circle problem in Navier-Stokes flow in two dimensions. In three-dimensional Navier-Stokes flow, which is by far the most important and also the most challenging of the cases that we have considered, the FSIB method shows second-order convergence with immersed boundaries that are topologically spherical and have mechanical properties derived from a surface elasticity model or from a surface tension model. We have no explanation for the different convergence rates observed in different settings, but in every case that we have tested that the FSIB method outperforms the standard IB method, either by having a higher order of convergence or a much smaller empirical error with the same order of convergence.  Our quantitative estimates of effective boundary thickness also show that the FSIB method has improved boundary resolution in comparison to the standard IB method. Because of these advantages, which are achieved without increased computational cost, we believe that the FSIB method will turn out to be widely applicable.

\section{Acknowledgement}

The first author, Zhe Chen, is supported by the Henry MacCracken Fellowship at New York University.

\begin{appendices}\label{sec:append}

\section{Nonuniform Fast Fourier Transform (NUFFT)}\label{appendix:nufft}

The NUFFT algorithm generalizes the FFT to the nonuniform data that is off the grids. Two types of NUFFTs are used in this paper and are called the type-1 NUFFT and the type-2 NUFFT. For simplicity, we denote the number of dimensions $d=3$ and set the number of desired Fourier modes $N$ in each spatial dimension to be even. 

\subsection{Type-1, nonuniform to uniform}

The goal of the type-1 NUFFT is to compute the nonuniform Fourier transform below efficiently and accurately. 

\begin{equation}\label{eq:type1}
\hat{f}_{\bs k}:=\sum_{j=1}^{N_b} f_j \exp(-i \bs k \cdot \bs x_j),\quad \bs k \in \mathcal{K}_{N}
\end{equation}

Note that definition of $\hat{f}$ is equivalent to Fourier transform of periodic function $f(\bs x)=\sum_{j=1}^{N_b} f_j \delta(\bs x- \bs x_j)$ in $[0,L]^3$. Here $f(\bs x)$ is a sum of the delta function with values $f_j$ in locations $x_j$. Also, one can consider this as a generalization of discrete Fourier transform(DFT). When $x_j$ happens to be on the uniform grids, the type-1 NUFFT becomes the DFT. We desire to compute only $N$ Fourier modes in $\mathcal{K}_{N}$ that is defined in \cref{sec:spatial_discretization}.

The first step is to spread the nonuniform source to the uniform grids $x_{i,j,j},\ i,j,k=1,...,N$ by convolution with a kernel function $G_\xi$ as

\begin{equation}\label{eq:type1_conv}
f_G(\bs x_{i,j,k}) := G_\xi * f (\bm{x_{i,j,k}})= \sum _{q=1}^{N_b} f_q G_\xi (\bm x_{i,j,k}-\bm x_q) .
\end{equation}

Due to the choice of $G_\xi$ that is discussed in \cref{sec:kern_choi}, we only need to spread to the nearest $\sigma$ grid points, i.e. the width of the kernel is truncated to $2\sigma h$. 

Then, we apply an FFT on the uniform grids to get the Fourier space and cancel the effect of the convolution with the kernel by dividing by the Fourier transform of the kernel, i.e. $\hat{G}_\xi$. The $\hat{G}_\xi$ is usually analytically available or efficiently computed. So, we could compute $\hat{f}$ by

\begin{equation}\label{eq:type1_deco}
    \hat{f}(\bm{k})\approx\frac{\mathcal{F} \left( f_G(\bs x_{i,j,k})\right)}{\hat{G}_\xi(\bs k)}, \quad \bm k \in \mathcal{K}_{N},
\end{equation}
where $\mathcal{F}$ denotes FFT.

\subsection{Type-2, uniform to nonuniform}

The type-2 NUFFT is usually considered as evaluating Fourier series on nonuniform points in physical space for given uniform Fourier modes in $\mathcal{K}_{N}$, which formulates to   

\begin{equation}\label{eq:type2}
f_j:=\sum_{\bs k \in \mathcal{K}_{N}} \hat{f}_{\bs k} \exp(i \bs k \cdot \bs x_j), \quad j=1,...,N_b.
\end{equation}
Therefore, $f_j$ is the value of the Fourier series at location $x_j$.

The algorithm of the type-2 NUFFT is like the reciprocal of the type-1. We first divide the $\hat{f}_{\bs k}$ by Fourier transform the same kernel, i.e. $\hat{G}_\xi$. Then, apply an inverse FFT to get function $g(\bs x_{i,j,k})$ on the grids as

\begin{equation}\label{eq:type2_divi}
g(\bs x_{i,j,k}) := \mathcal{F}^{-1}\left(\frac{\hat{f}}{\hat{G}_\xi}\right) .
\end{equation}

Last, spread it to the nonuniform off-grid points $x_q$ by the kernel $G_\xi$ as

\begin{equation}\label{eq:type2_conv}
f_q=\sum_{i,j,k} g(\bs x_{i,j,k})  G_\xi(\bs x_q- \bs x_{i,j,k})h^3,
\end{equation}
where $q=1,...,N_b$ and $\mathcal{F}^{-1}(\cdot)$ denotes the inverse FFT. This cancels the effect of $G_\xi$ in \cref{eq:type2_conv} by the convolution theorem. Again, the spreading is truncated to the nearest $\sigma$ grid points.

\subsection{Choice of the kernels and its parameters}\label{sec:kern_choi}

For the kernel $G_\xi$ to be good, the criteria are simple. First, $G_\xi$ should be able to be truncated to as small support as possible, i.e. $G_\xi$ decays fast in physical space. Second, the tail of $\hat{G}_\xi$ should be as small as possible relative to the truncation windows $|k_j|<N/2$. Third, $\hat{G}_\xi$ should be easily available. Here we provide a popular choice of $G_\xi$, the "Kaiser-Bessel" kernel $G_{\text{KB},\xi}$ \citep{barnettParallelNonuniformFast2019} 

\begin{equation}\label{eq:kernel_KB}
G_{\text{KB}, \xi}(x):=\left\{\begin{array}{ll}
I_{0}\left(\xi \sqrt{1-x^{2}}\right) / I_{0}(\xi), & |x| \leq 1 ,\\
0, & \text { otherwise },
\end{array}\right.    
\end{equation}
where $I_0$ is the regular modified Bessel function of order zero. Its Fourier transform $\hat{G}_{\text{KB},\xi}$ is analytically available as

\begin{equation}\label{eq:kernel_KB_hat}
\hat{G}_{\text{KB}, \xi}(k)=\frac{2}{I_{0}(\xi)} \frac{\sinh \sqrt{\xi^{2}-k^{2}}}{\sqrt{\xi^{2}-k^{2}}}.
\end{equation}

Because $G_\xi$ decays exponentially in physical space, we choose truncation in physical space $\sigma \sim |\log(\epsilon)|$ for any desired accuracy $\epsilon$ of \cref{eq:type1} and \cref{eq:type2} compared with exact nonuniform Fourier transform. Then, we should choose the parameter $\xi$ such that the tail of $\hat{G}_\xi$ has a smaller error than $\epsilon$. One may refer to \citep{barnettParallelNonuniformFast2019} for further details on error analysis and parameter choice. 

As for the computational cost, the direct computation of the nonuniform Fourier transform is $O(N_b\cdot N^3)$. Considering $N_b$ is proportional to $N^2$ in three dimensions, this is an unbearable computational cost. By using NUFFT, we reduce the cost to $O(N_b\sigma^3+N^3\log(N))$, the first term of which is the spreading cost and the second is the FFT's cost.

\end{appendices}

\bibliographystyle{plain} 
\bibliography{references, refs}

\end{document}